\documentclass[letterpaper]{emulateapj}
\usepackage{multirow}
\usepackage{graphics}
\usepackage{amsmath, amsthm, amssymb}

\newcommand{\sech}{{\rm sech}}

\begin{document}

\title[The Effect of Star Formation and FUV on GMCs]{Star Formation in Disk Galaxies. II. The Effect of Star
  Formation and Photoelectric Heating on the Formation and Evolution
  of Giant Molecular Clouds}

\author{Elizabeth J. Tasker\altaffilmark{1,2,3}}
\altaffiltext{1}{CITA National Fellow} 
\altaffiltext{2}{Department of Physics and Astronomy, McMaster University, 1280 Main Street West, Hamilton, Ontario L8S 4M1, Canada.} 
\altaffiltext{3}{Department of Astronomy, University of Florida, Gainesville, FL 32611, USA}

\begin{abstract}
We investigate the effect of star formation and diffuse photoelectric
heating on the properties of giant molecular clouds (GMCs) formed in high
resolution ($\lesssim 10$\,pc) global ($\sim 20$\,kpc) simulations of
isolated Milky Way-type galaxy disks. The clouds are formed through
gravitational fragmentation and structures with densities $n_{\rm H, c} >
100$\,cm$^{-3}$ are identified as GMCs. Between 1000-1500 clouds are
created in the simulations with masses $M > 10^5$\,M$_\odot$ and 180-240 with masses $M >
10^6$\,M$_\odot$ in agreement with estimates of the Milky Way's
population. We find that the effect of
photoelectric heating is to suppress the fragmentation of the ISM,
resulting in a filamentary structure in the warm gas surrounding
clouds. This environment suppresses the formation of a retrograde
rotating cloud population, with 88\% of the clouds rotating prograde
with respect to the galaxy after 300\,Myr. The diffuse heating also
reduces the initial star formation rate, slowing the conversation of
gas into stars. We therefore conclude that the interstellar
environment plays an important role in the GMCs evolution. Our clouds
live between $0 - 20$\,Myr with a high infant mortality ($t' <
3$\,Myr) due to cloud mergers and star formation. Other properties,
including distributions of mass, size and surface density agree well with
observations. Collisions between our clouds are common, occurring at a
rate of $\sim 1/4$ of the orbital period. It is not clear whether such
collisions trigger or suppress star formation at our current
resolution. Our star formation rate is a factor of 10
higher than observations in local galaxies. This is likely due to the
absence of localized feedback in our models. 
\end{abstract}


\maketitle

\section{Introduction}

Embedded in the turbulent gas of the interstellar medium (ISM) are the
stellar nurseries of the Galaxy; the giant molecular clouds
(GMCs). The properties of these clouds dictate the environment in
which most new stars are formed, making these objects the primary
controllers of star formation in the Galaxy. To understand star
formation, it is therefore essential that we first understand the
processes that govern the formation and evolution of the GMCs.

Exactly what determines the GMC properties is a taxing
problem. Theoretical work on their formation has led to two different
schools of thought. The ``Top-down'' mechanism suggests GMC formation
is driven via large-scale gravitational or magnetic disk instabilities
\citep[e.g.][]{Shetty2006, Kim2003, Glover2007a, Glover2007b}, whereas
``bottom-up'' processes see clouds formed in colliding flows
\citep[e.g.][]{Heitsch2008} or via agglomeration from inelastic
collisions between the GMCs \citep{Kwan1979}. It is possible that both
these processes are important in different galactic environments
\citep{Dobbs2008}. 

Depending on the lifetime of the cloud, the importance of the
formation mechanism on the GMC properties varies. If the cloud lives
as long as a few free-fall times, then its attributes could depend
primarily on interactions with its environment (e.g. disk sheer,
spiral arms and cloud-cloud collisions), and also on its internal
processes driven by star formation, including turbulence from
supernovae and radiation pressure. Whether GMCs are long-lived enough
for this is highly debated with theories supporting lifetimes both
short and long compared to the free-fall timescale, 

\begin{equation}
t_{ff} = \left(\frac{3\pi}{32 G \rho}\right)^{1/2} = 4.35\times
10^6\left(\frac{n_{\rm H}}{100 {\rm cm}^{-3}}\right)^{-1/2} {\rm yr.}
\end{equation}

However, current estimates suggest that a GMC lives between 1-2
free-fall times \citep[e.g.][and references therein]{McKee2007}. This
implies that the cloud's evolution could play a significant role in
determining its properties. 

Observationally, it is difficult to measure the cloud properties
accurately since the molecular gas cannot be measured directly, but
must be inferred from the abundances of CO \citep{Glover2010,
  Shetty2010}. From the theoretical stand point, there is the
problem of scale that the forces are operating on. While the GMC itself
is of order 20\,pc in radius, it lives in a galactic disk whose
stellar component, in the Milky Way, extends to around 20\,kpc in
radius. This means to self-consistently replicate their properties,
simulations of GMCs must encompass 3-4 orders of magnitude in
scale. Previous work that has studied the properties of GMCs on the
galactic scale has often been limited to two dimensions
\citep{Shetty2008} or has had to assume a fixed two-phase medium for the
ISM \citep{Dobbs2008}. While the majority of the clouds are confined
to the plane of the disk, both cloud collisions and feedback eject gas
from the surface and, in the latter instance, it is this vertical
expulsion that is thought to regulate the pressure in the ISM
\citep{McKee1977, Cox2005}; the environment in which the
clouds are forming. Similarly, previous three dimensional
studies of the ISM in galaxy disks at lower resolution, produce a
continuous multiphase medium that is poorly represented by a
fixed, discrete phase model \citep{Tasker2006, Tasker2008, Wada2007}.

Local models which consider a partial region of the galaxy disk (normally less
than a few kpc across) are able to achieve much higher resolutions and
--with the use of a shearing box-- can approximate the effects of
galactic shear on the cloud properties \citep{Kim2008, Kim2007, Kim2006,
  Kim2001}. Alternative models on similar scales investigate the
results from colliding flows \citep{Heitsch2008, Heitsch2009} and the
local impact of supernovae and turbulence \citep[e.g.][]{Joung2006,
  Slyz2005}. These simulations are able to tell us much about the
structure of the GMCs and star-forming gas, but
they cannot explore the evolution of the GMCs as they move
through the disk or compare with globally averaged properties. 

Whether the global environment is important for correctly modeling
the evolution of the GMCs remains an open question. If GMC properties
are not a strong function of their
environment, then they can be modeled as separate entities and the
inclusion of global forces from the galactic disk
are not necessary. This hypothesis is given some support by the
observational results from populations of GMCs in galaxies other than
the Milky Way, which find their properties are similar to those of Galactic GMCs, including
their velocity dispersions, surface density and virial parameters
\citep{Bolatto2008, Rosolowsky2003, Fukui2008}. \citet{Bigiel2008}
also found that the 18 galaxies in The HI Nearby Galaxy Survey (THINGS)
appeared to have a fixed star formation rate (SFR) per unit of molecular gas,
all of which could suggest a set of universal properties for GMCs.  

On the other hand, if the clouds are gravitationally bound, then their
velocity dispersions become established via GMC interactions
\citep{Gammie1991}. In this case, their radial position in the galaxy
can change, causing greater susceptibility to galactic shear. Both
\citet{Gammie1991} and \citet{Tan2000} argue that self-gravitating
GMCs should suffer relatively frequent collisions, which could be an
important process in controlling the molecular mass of a GMC. This
would impact the GMC's SFR, making the
galactic environment an essential ingredient in the GMC's
evolution. Furthermore, while the mass profile of GMCs is
universally seen to be a power law of the form,

\begin{equation}
\frac{dN_c}{d\ln M_c}\propto M_c^{-\alpha_c},
\end{equation}

the value of the exponent, $\alpha_c$, is
found to differ between galaxies. \citet{Williams1997} measure a value of
$\alpha_c$ between $0.6 -0.8$ in the Milky Way, while
\citet{Rosolowsky2003} finds a steeper gradient of $\alpha_c \simeq 1.6$
in M33. \citet{Blitz2004} conclude that these variations are not due
to systematic uncertainties, but are a consequence of the galactic
environment. This suggests that the global-scale structures impact the range of
masses of clouds formed. 

Perhaps most suggestive evidence in favor of an intrinsic connection between
clouds and their galactic environment is the Kennicutt-Schmidt
relation \citep{Kennicutt1998}. This empirical relation links the
averaged gas surface density, $\bar{\Sigma}_{\rm gas}$ to the surface
density of star formation rate $\bar{\Sigma}_{\rm sfr}$ via:

\begin{equation}
\bar{\Sigma}_{\rm sfr} \propto \bar{\Sigma}_{\rm g}^{\alpha_{\rm sfr}}
\label{eq:ks}
\end{equation}

\noindent where \citet{Kennicutt1998} found the exponent $\alpha_{\rm sfr}
= 1.4\pm 0.15$ for spatial averaging over the entire disk. More recent
studies using the THINGS data
\citep{Bigiel2008} find the relation to hold linearly with the molecular
component of the gas with a $\alpha_{\rm sfr} = 1.0\pm 0.2$, with
spatial averaging (resolution) of 750\,pc. 

An alternatively fit for this data can be found by relating 
$\bar{\Sigma}_{\rm sfr}$ to the orbital angular frequency at the outer
radius, $\Omega_{\rm out}$: 
$\bar{\Sigma}_{\rm sfr} \propto \bar{\Sigma_g}\Omega_{\rm out}$. This
result applies even in the more extreme environments of starburst
and high red-shift galaxies \citep{Genzel2010}. Both
these relations indicate an intimate connection between global
structure and the GMC star-forming environment.

On scales of the same order as the GMC size, an important
component in the GMC evolution must come from the stars
themselves. Star formation is observed to be highly clustered, with
star clusters forming out of dense clumps with initial radii $\sim 1$\,pc
\citep{Lada2003}. Within this small region, the total star formation efficiency is
relatively high at $\sim 0.1-0.5$, but the majority of the GMC is not forming
stars, possibly due to the effect of magnetic fields
\citep{Crutcher2005}. The average efficiency over the whole cloud is
therefore of order a few percent per local free-fall time \citep{Krumholz2007, Zuckerman1974}. 

Gas is removed from the cloud to create a star which then deposits
energy into its surrounding medium through diffusive
and energetic feedback. During its lifetime, a massive star will be source of FUV
radiation which can be absorbed by dust grains to eject an electron
that will heat the gas. This photoelectric heating has long been
thought to be the dominant form of heating in the neutral ISM, which
includes the GMC population \citep{Wolfire1995}. More energetic
forms of feedback from stellar winds and supernovae will also act to
deposit concentrated blasts of energy into the star's immediate
environment. Whether the cloud can survive the star's
life-cycle is debated \citep[e.g.][]{Murray2010} but the fact this
process will affect the cloud's properties is not.

Due to the range of the forces in play, it is exceptionally difficult to
determine the dominant processes affecting a GMC's evolution. Is it
the interaction with the global galaxy environment, the results of
star formation and feedback or an equal combination of systems? In these
papers, we aim to investigate this with a set of global disk
simulations that separate out the processes by introducing each influencing factor individually. 

In our first paper \citep[hereafter TT09]{Tasker2009}, we simulated an
idealized population of GMCs without the presence of star formation or
any form of feedback. Despite the simplicity of the model, we
reproduced many of the observable properties of measured GMC
populations including mass surface density, velocity dispersion,
angular momentum and vertical distribution. In addition to this, we
found a typical collision time between clouds of $\sim 20$\,\% of the
local orbital time, in agreement with estimates
by \citet{Tan2000}. This suggested that compressive flows generated in
cloud collisions could be a dominant mechanism for inducing star
formation. Such a process has parallels with the local-scale colliding flow models
which trigger star formation via compressive forcing of the turbulent
flows \citep{Banerjee2009, Heitsch2008, Hennebelle2008}. However, for this
simulation without star formation and feedback processes
included in the model, the GMCs
could only be destroyed through mergers. This meant that it was not
possible to regulate their evolution, giving a steadily more massive
population over time.

In this paper, we extend the model in two new simulations. The first
of these includes star formation without feedback and the second
contains star formation with diffuse feedback from FUV
photoelectric heating. We compare the properties of the GMC
populations formed in both models and with the GMCs formed without star
formation in TT09. Our results will show that the diffuse heating
reduces the fragmentation of the disk, causing clouds to be embedded
in a filamentary warm ISM. This reduces the initial
star formation rate and suppresses the formation of a retrograde
rotating population of clouds. We will see that both our population
of clouds continue to match many of the observations of GMCs in the Milky Way,
including the mass profile, size, mass weighted surface density,
vertical distribution and gravitational binding. With the inclusion of
cloud destruction through star formation and mergers, we are able to estimate the
average lifetimes of the clouds in our populations which 
are found to be largely between $0-20$\,Myr in agreement with current
estimates. We will show that our SFR is a factor of 10 higher than
that observed in local galaxies and suggest that this is due to the
lack of energetic local feedback in our models. 

These localized feedback processes, such as supernovae, radiation pressure
and ionization are expected to play an important role in GMC
evolution. However, to understand the determining forces on the GMC
population, we focus on the two effects of star formation and diffuse
heating in this paper, leaving additional forces for later study. 

Details of our method are outlined in section\,\S 2, global properties of the
disk are presented in \S 3 and the structure of the ISM in \S 4. \S 5
will focus on the properties of the individual GMCs and \S 6 will look
at the star formation in the disk. In \S 7 we will present our
conclusions. \\*[0.2cm]

\section{Numerical Techniques}

\subsection{The Code}
\label{sec:code}

The simulations presented in this paper were run using {\it Enzo}; a
three-dimensional adaptive mesh refinement (AMR) hydrodynamics code
\citep{OShea2004, Bryan1999, Bryan1997}. The AMR technique is
particularly adept at resolving multiphase fluids such as those found
in the ISM, due to the natural boundaries between grid cells producing
accurate resolution of shocks and low numerical mixing
\citep{Tasker2008b}. {\it Enzo} has previously been used to
model galactic disks where it successfully produced a self-consistent
atomic multiphase ISM \citep{Tonnesen2010, Tasker2009, Tasker2008, Tasker2006}. 

We used a three-dimensional box of side $32$\,kpc with a root grid of
$256^3$ and four levels of refinement, giving a limiting resolution (smallest
cell size) of $7.8$\,pc. For the effect of resolution on our
simulation results, see TT09 where a detailed study is presented.
Gas was refined whenever the Jeans' Length was resolved by less than
four cell widths, as suggested by \citet{Truelove1997} as the
resolution needed for avoiding artificial fragmentation. On our finest
level, the Truelove criteria is
maintained until $\sim 100$\,cm$^{-3}$, our threshold density for
cloud definition. A discussion on resolving the gravitational collapse
in the simulation is presented more fully in TT09.

The evolution of the gas in {\it Enzo} was performed using a
three-dimensional version of the {\it Zeus} hydrodynamics algorithm
\citep{Stone1992}. This routine uses an artificial viscosity term to
represent shocks, where the variable associated with this, the quadratic
artificial viscosity, was set to 2.0 (the default) for all
simulations. 

Radiative cooling was included using rates from the analytical
expression of \citet{Sarazin1987} for solar metallicity to $10^4$\,K
and down to $T=300\,K$ using rates from \citet{Rosen1995}. This allows the
gas to cool to the temperature of the upper end of the atomic cold
neutral medium \citep{Wolfire2003}. Actual GMCs will have temperatures
of $\sim 10$\,K, an order of magnitude below our minimum
temperature. However, at our resolution, clouds with
diameters of 100\,pc only have 13 cells in each linear dimension (with
an average GMC having a radius of 16\,kpc and 4 cells across), which is
insufficient to resolve the full turbulent structure of the gas. We
also do not include pressure from magnetic fields, so imposing this
temperature floor of 300\,K produces a minimum sound speed of $1.8$\,kms$^{-1}$
to crudely allow for these effects. In fact, the velocity dispersion within
our clouds is typically higher than this by about a factor of two, implying that this
floor is not having a significant impact on our cloud properties. 

In addition to radiative cooling, the gas can also be heated via
diffuse photoelectric heating in which electrons are ejected from dust
grains via FUV photos. In the simulation where this was turned on, we
included a radially dependent heating term of the form described in \citet{Wolfire2003}:

\begin{equation}
\label{eq:peheat}
\Gamma_{\rm pe} = 10^{-24} \epsilon_h G_0 \left\{ 
\begin{array}{ll}
e^{-(R-R_0)/H_R} \, {\rm ergs\, s^{-1}} & \textrm{r $\ge 4.0$\,kpc} \\
e^{-(4-R_0)/H_R} \, {\rm ergs\, s^{-1}} & \textrm{r $< 4.0$\,kpc}
\end{array} \right.
\end{equation}\\

\noindent where the heating efficiency $\epsilon_h = 0.05$ and $G_0$ is
the incident far-ultraviolet field normalized to the \citet{Habing1968} estimate
for the local ISM value. We take a value of $G_0 = 1.7$ in agreement
with \citet{Draine1978}. $R_0$ is the radial scale length at 8.0\,kpc and
$H_R = 4.1$\,kpc, the scale length as estimated by \citet{Wolfire2003}.

Collisionless star particles, representing star clusters,  are allowed
to form in our simulation in the main region
of the disk between $2.5 < r < 8.5$\,kpc. As is described in
\ref{sec:ic} below, this is the area of our disk where we identify
and analyze the GMCs. Within this region, star particles are created
when the density within a cell exceeds the threshold value of
$n_{\rm H} = 100$\,cm$^{-3}$. Since our finest refinement cells (where
the stars will be formed) are still 7.8\,pc across, the gas within
them can be assumed to be turbulent. We therefore do not check for
gravitational collapse or boundedness of the cell gas, since such processes
are likely to affect star formation on much smaller scales than what
we can resolve. Cells with temperatures greater
than 3000\,K are also prevented from forming stars to rule out the
possibility of star formation in the hot dense gas of shock
fronts. In practice, star particles form in gas typically close to
300\,K, so this limit is not particularly important. When a cell
reaches the threshold density, a star particle is
created whose mass is calculated by:

\begin{equation}
m_* = \epsilon_{\rm ff}\frac{\Delta t}{t_{\rm ff}} \rho_{\rm gas} \Delta x^3
\label{eq:starmass}
\end{equation}

\noindent where $\epsilon$ is the star formation efficiency (the
fraction of gas that is converted into star particles per dynamical time),
$\Delta t$ is the size of the time step, $t_{\rm ff}$ is the time for
dynamical collapse in the cell ($t_{\rm ff} =
\left(\frac{3\pi}{32G\rho}\right)^{1/2}$) and $\rho_{\rm gas}$ is the gas
density. The resultant object should be considered a star cluster,
rather than an individual star since it contains approximately
1000\,M$_\odot$. For our simulations, we chose a value
for $\epsilon_{\rm ff} = 0.02$, in agreement with the observational
constraints described by \citet{Zuckerman1974} for GMCs and
\citet{Krumholz2005} for GMCs and their internal, higher density components. 

An additional computational requirement is that the star particle
will not be created if the calculated value for $m_*$ is less than a
given minimum value of m$_{\rm min} = 10^3$\,M$_\odot$. This purely
numerical addition is to prevent the calculation becoming prohibitively
slow due to a extremely large number of star particles. In the
situation where this occurs, an override exists whereby a particle of
mass $m_{\rm min}$ is created with a probability equal to the ratio
between the mass of the would-be star particle and $m_{\rm min}$. In
practice, almost all star particles born in the
highest resolution simulations are created via this stochastic method
with masses equal to $m_{\rm min}$.

The motions of the star particles are calculated as a collisionless
N-body system. They interact gravitationally with the gas via a
cloud-in-cell mapping of their positions onto the grid to produce a
discretized density field. The number of star particles created during
the simulations is 2.5 - 3 million. In this paper, there is no
localized feedback to the gas directly from the star particle. 

\subsubsection{Runs Performed}

This paper presents the results from two different simulations, both at a
limiting resolution of 7.8\,pc.

The first simulation (simulation {\it disk SFOnly}) includes star formation, implemented as described above,
but with no form of additional heating. The second simulation
(simulation {\it disk SF+PEheat}) also includes star formation and a radially
dependent diffuse heating term with the form given by Equation~\ref{eq:peheat}.

The results in the absence of both star formation and diffuse heating
are presented in TT09, simulation {\it disk NoSF}.

\subsection{Galaxy Initial Conditions and Cloud Analysis\\*[0.2cm]}
\label{sec:ic}

The initial conditions for the simulations are described in detail in TT09. They
consist of an isolated gas disk sitting in a static background
potential that represents both a dark matter halo and a stellar disk
component for a galaxy similar to the Milky Way. The potential gives
the disk a constant circular velocity for $r >> 0.5$\,kpc of $v_c = 200$\,km\,s$^{-1}$. 

We focus on the gas in the main region of the disk between $2.0 < r <
10.0$\,kpc. Gas here is initially marginally stable against
gravitational collapse, with the Toomre Q parameter for gravitational
instability \citep{Toomre1964} having a constant value:

\begin{equation}
Q = \frac{\kappa\sigma_g}{\pi G \Sigma_g} \sim 1.5
\label{eq:toomre}
\end{equation}

\noindent where $\kappa$ is the epicycle frequency and $\Sigma_g$, the
gas surface density. $\sigma_g \equiv \sqrt{\sigma_{nt}^2 + c_s^2}$ is
the mass-weighted 1D velocity dispersion of the gas, with
$\sigma_{nt}$ being the velocity motions in the disk plane after
subtraction of the circular velocity. The
vertical profile is proportional to $\sech^2(z/z_h)$, where the scale
height, $z_h$ varies with galactocentric radius based on HI
observations of the Milky Way \citep{BM1998}. At the solar radius of
8\,kpc, $z_h = 290$\,pc. For a flat rotation curve, $\kappa =
\sqrt{2}v_c/r$, which gives a gas density profile of the form:

\begin{displaymath}
\rho(r,z) = \left(\frac{\sqrt{2}v_c\sigma_g}{4\pi G Q z_h}\right)\frac{1}{r}\sech^2\left(\frac{z}{z_h}\right)
\end{displaymath}

\noindent Low density regions of gravitationally stable gas sit in the disk center between
$0 < r < 2$\,kpc and at the outer edge between $10 < r < 12$\,kpc. 

As the gas cools, the main region of the disk becomes gravitationally
unstable with $Q < 1$ and fragments. Star formation and cloud analysis
are restricted to between $2.5 > r > 8.5$\,kpc; within the main region
but avoiding boundary effects.

Details of the algorithm used to identify and track the GMCs in the
simulations are described in TT09. In short, we identify ``GMCs'',
i.e. star-forming clouds, as peaked and coherent structures contained
within contours of the threshold density of $n_{\rm H,c} \ge 100$\,cm$^{-3}$,
about the mean volume density of observed galactic GMCs. It is
worth remembering that while GMCs are by definition molecular, our gas
is purely atomic, so we are selecting structures that would be mostly
molecular in reality, although our procedure does not distinguish
dense atomic gas that might be present in photodissociation regions. By comparing outputs
of the simulation at 1\,Myr intervals, the clouds are tracked over the
course of the simulation to produce a timeline of their evolution. The
simulations themselves are run for 300\,Myr, just over one orbital
period at the outer edge of the main region in the disk.

\section{Global Evolution of the Galactic Disk}
\label{sec:globalevol}

\begin{figure*}
\begin{center}
\includegraphics[width=\textwidth]{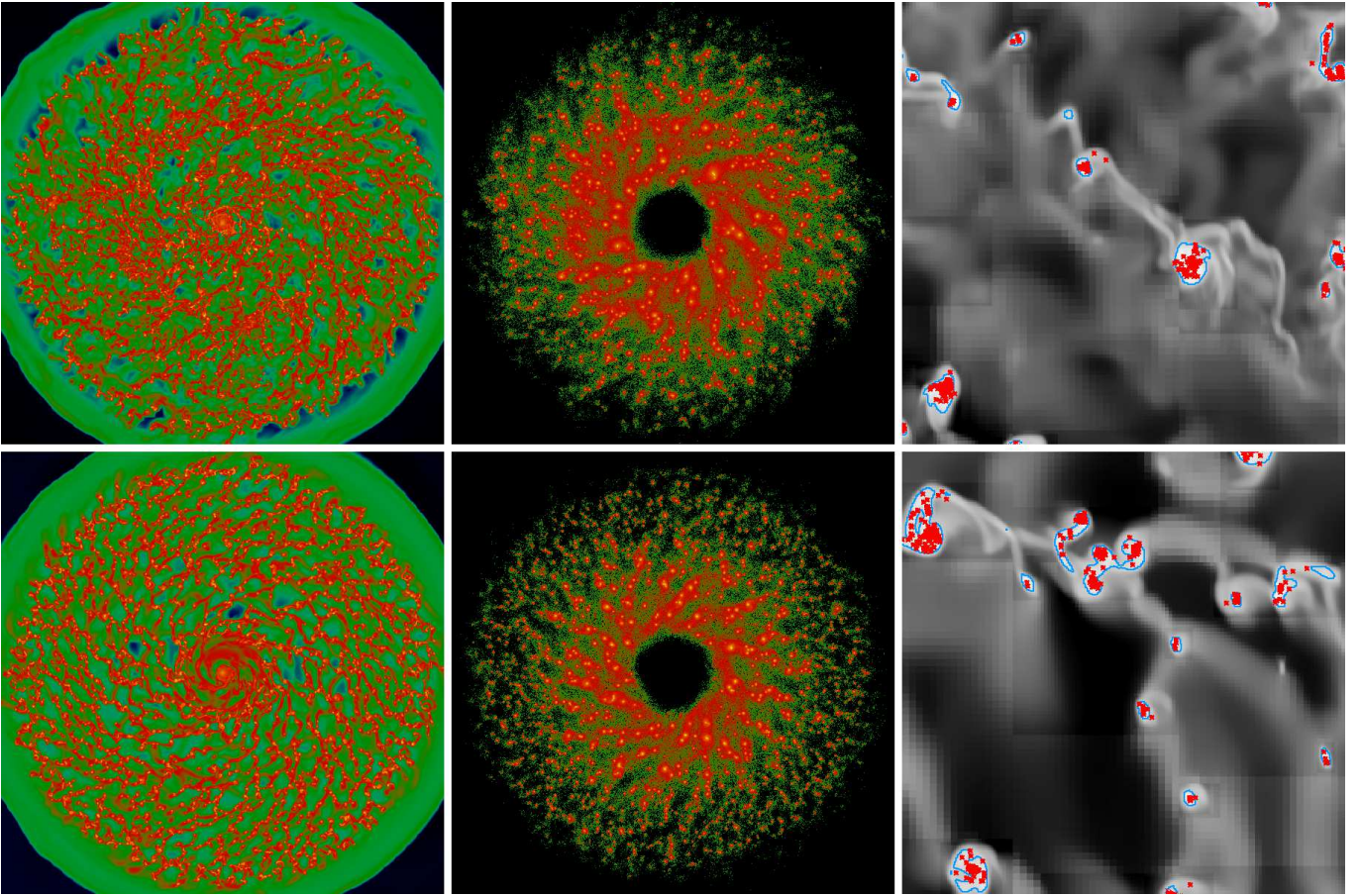}
\caption{Images of the galactic disk at 200\,Myr. Top panel is for
  disk SFOnly while the bottom panel shows disk SF+PEheat. The left-hand image is 20\,kpc
  across and shows the log-scaled surface density of the disk with range $[0.0033,
  3715]$\,M$_\odot pc^{-2}$. The center pane is the projected star
  particle density and the right-hand image shows a 2\,kpc density
  slice of the mid-plane. Blue contour lines mark the cloud boundaries
  corresponding to a number density of $100\,$cm$^{-3}$ and the new
  star particles (age $< 1$\,Myr) are shown in red.
\label{fig:diskimages}}
\end{center}
\end{figure*}

\begin{figure*}
\begin{center}
\includegraphics[angle=270,width=\textwidth]{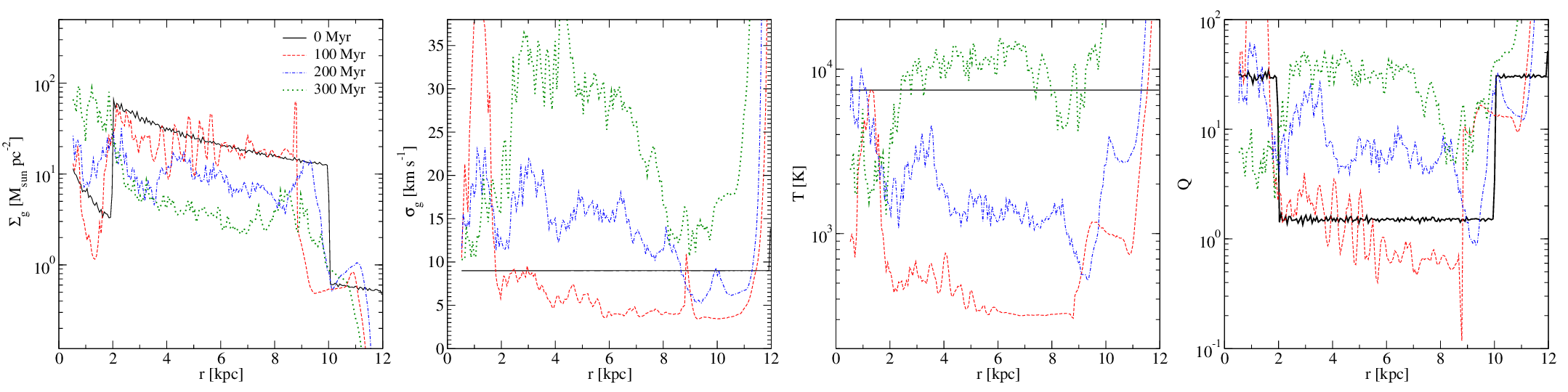}
\includegraphics[angle=270,width=\textwidth]{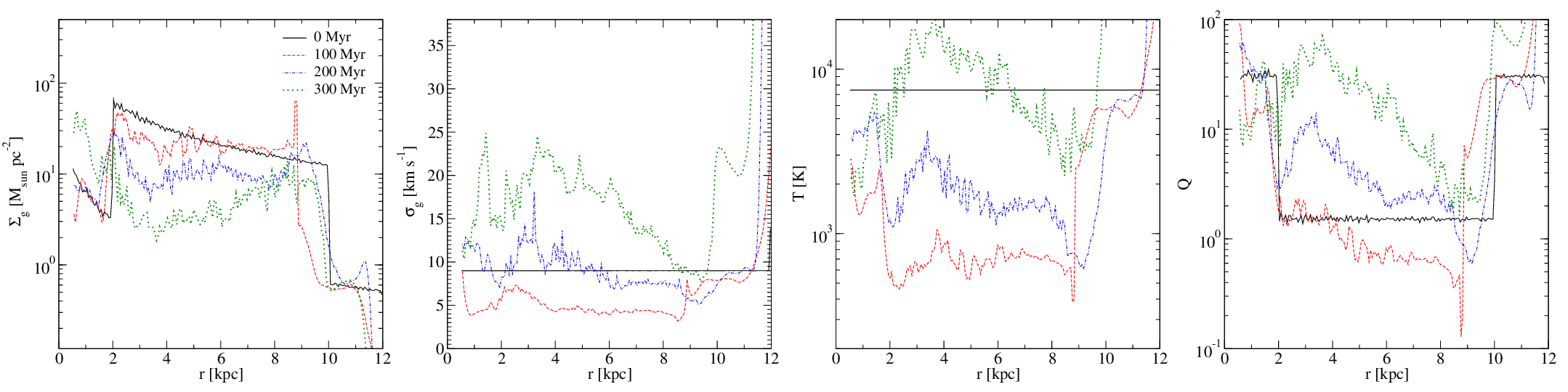}
\caption{Evolution of the azimuthally-averaged radial galactic profiles
  for disk SFOnly (top row) and disk SF+PEheat which (bottom row). Plots left to right
  show: (a) gas mass surface density ($\Sigma_g$), (b) 1D velocity dispersion of
  the gas ($\sigma_g$), (c) temperature ($T$) and (d) Toomre $Q$
  parameter. Note, $\Sigma_g = \int_{-1{\rm kpc}}^{+1{\rm kpc}}\rho(z) d z$,  $T$ is a
  mass-weighted average over $-1~{\rm kpc}<z<1~{\rm kpc}$ and $Q$
  makes use of $\sigma_g$ evaluated as a mass-weighted average over the same volume. 
\label{fig:diskevol}}
\end{center}
\end{figure*}

The disk rapidly cools from its initial conditions to fragment through
gravitational instabilities (see TT09 for a discussion of this process). By $t = 140$\,Myr, the
main region of the disk between $2.5 > r > 8.5$\,kpc has fully
fragmented into dense cold clouds of gas, embedded in a warmer medium.

Three images of the disk at $t = 200$\,Myr are shown in
Figure~\ref{fig:diskimages} for simulation disk SFOnly (top row) and
disk SF+PEheat (bottom row), where diffuse heating is included. The
left-hand panel shows the gas surface density over the main region of
the disk, the middle panel shows the star particle density for the
same region and the right-hand panel displays a $2$\,kpc slice of a typical
patch in the galactic mid-plane. In this right-hand panel, blue
contours mark our cloud threshold definition of $100$\,cm$^{-3}$ and
new star particles, with age $< 1$\,Myr, are shown in red.  

The effect of including the diffuse heating on the disk's ISM can be seen in
the images of the gas surface density. Without the additional heating
in disk SFOnly,
the gas collapses to form smaller structures, especially in the inner
region, $2 > r > 4$, where both the density of the disk is highest
(and hence the dynamical time shortest) and the heating term, as
described in equation~\ref{eq:peheat}, is at its maximum. While
the disk has evidently become gravitationally unstable in both cases, its
fragmentation is reduced by the diffuse heating.

One of the main results of the reduced fragmentation is shown in the
middle panel of Figure~\ref{fig:diskimages} in the distribution of
star particles. The stellar density is visibly lower when diffuse heating is
included. At $t = 200$\,Myr, disk SFOnly has
formed over 3 million star particles, whereas this is reduced to 2.6 million
in disk SF+PEheat. The star formation properties of the
disk will be considered quantitatively in section~\ref{sec:stars}.

The findings from the first two image panes in
Figure~\ref{fig:diskimages} are supported in the
close-up of the $2$\,kpc slice of the galactic plane on the
far-right. The filaments connecting sites of star formation are denser
and thicker in disk SF+PEheat and the smaller star formation regions appear to have less
stars within them than ones of similar diameter in disk SFOnly.

The new star particles are born inside the identified cloud boundaries, as
expected since the density threshold for both cloud identification and star
formation is $n_{\rm H, c} = 100$\,cm$^{-3}$. In the occasional
case where a star particle appears to exist just outside the cloud
boundary, it has either moved during the last
1\,Myr or destroyed (via mass removal) the small part of its cloud
that it was formed in. With diffuse heating, the gas
between the clouds forms thicker filaments, supported against further
fragmentation by the thermal addition. This gives the warm ISM
between the clouds a more coherent structure.  

This variation in the global structure of the disk, together with its
evolution, is shown quantitatively in Figure~\ref{fig:diskevol}. The
four plots show azimuthally averaged radial profiles of the disk
properties for disk SFOnly (top row) and disk SF+PEheat (bottom). From
left to right, the plots are: (1) gas surface density,
$\Sigma_g = \int^{+1\,kpc}_{-1\,kpc}\rho(z)dz$, (2) the gas velocity
dispersion, $\sigma_g$, calculated via a mass-weighted average over
$-1\,{\rm kpc} < z < 1$\,kpc utilizing only disk plane velocity components,
(3) the mass-weighted temperature, $T$, averaged over the same range
as $\sigma_g$ and (4) the Toomre Q parameter for gravitational
instability as given in Equation~\ref{eq:toomre}. The four lines on
each plot show the profile at different times (t = 0, 100, 200 and
300\,Myr) during the simulation.

The initial conditions at $t = 0$ are shown by the black solid line in all
plots. We can see our main disk region between $2.5 < r < 8.5$\,kpc
is initially at a surface density higher than the surrounding gas by
over an order of magnitude and has a constant borderline stable $Q =
1.5$. The initial temperature is constant over the entire simulation
box and corresponds to a sound speed of $c_s = 9$\,kms$^{-1}$.

During the first 100\,Myr (red dashed line) the gas cools, bringing $Q$
below the critical stability value of 1.0 and causing the gas to
fragment. In the top panel showing disk SFOnly, the temperature drops until it approaches
the floor of our radiative cooling curve at $300$\,K, whereas the
addition of diffuse heating in the bottom panel slows down the cooling
of the densest clumps, increases the average minimum
temperature reached 
by approximately a factor of 2.0. Global ring instabilities appear in the
surface density profiles of both simulations (see TT09 for images of
their formation), but are less prominent in the gas warmed by the
diffuse heating, which provides an extra thermal support. As the disk fragments
tangentially, these fluctuations flatten out. The magnitude of the
surface density remains almost constant over the disk's main region during this
early period. At lower radii, however, the circular motion becomes
poorly resolved by the Cartesian grid, causing an infall of low density gas from the inner
region of the disk to build up at the disk center. 

By the time the simulation reaches 200\,Myr (blue dot-dashed line),
star formation has depleted the gas, causing the surface density to
drop by a factor of 2.0 in disk SFOnly and 1.5 for disk SF+PEheat. Since
star formation occurs in the coldest and densest regions of the disk,
the azimuthally averaged temperature rises as do the corresponding
values for the velocity dispersion and $Q$. The velocity dispersion
also increases due to gravitational heating, as star particles are formed to
create local deep potential wells. This effect is strongest in the disk
without diffuse heating, since 500,000 more star particles have formed by this
time, causing a greater depletion of the dense gas. 

Star formation continues to deplete the gas, resulting in the surface
density dropping by a factor of about 6 from the initial conditions by
300\,Myr at $r = 6$\,kpc for both runs (green dotted line). The SFR is
highest in the densest gas (since it is proportional to the free-fall
time), causing the surface density to drop first in the inner part of
our main region, $2.5 < r < 4$\,kpc. This is marked most in disk
SF+PEheat, where the star formation progresses more slowly due to the
reduced fragmentation.

The velocity dispersion is significantly lower in disk SF+PEheat,
having increased more slowly over the 300\,Myr. This
initially seems surprising, since the temperature is higher, but diffuse
heating has reduced the ability of the disk to fragment into massive
bound clusters, as was seen in
Figure~\ref{fig:diskimages}, leaving a filamentary state that reduces
the in-plane velocity motions, $\sigma_{\rm nt}$, (and the
corresponding $Q$ value),
that are excited by gravitational interactions between fragmented clumps. 

\section{The structure of the ISM}
\label{sec:ism}

\begin{figure*} 
\begin{center} 
\includegraphics[angle=270, width=\textwidth]{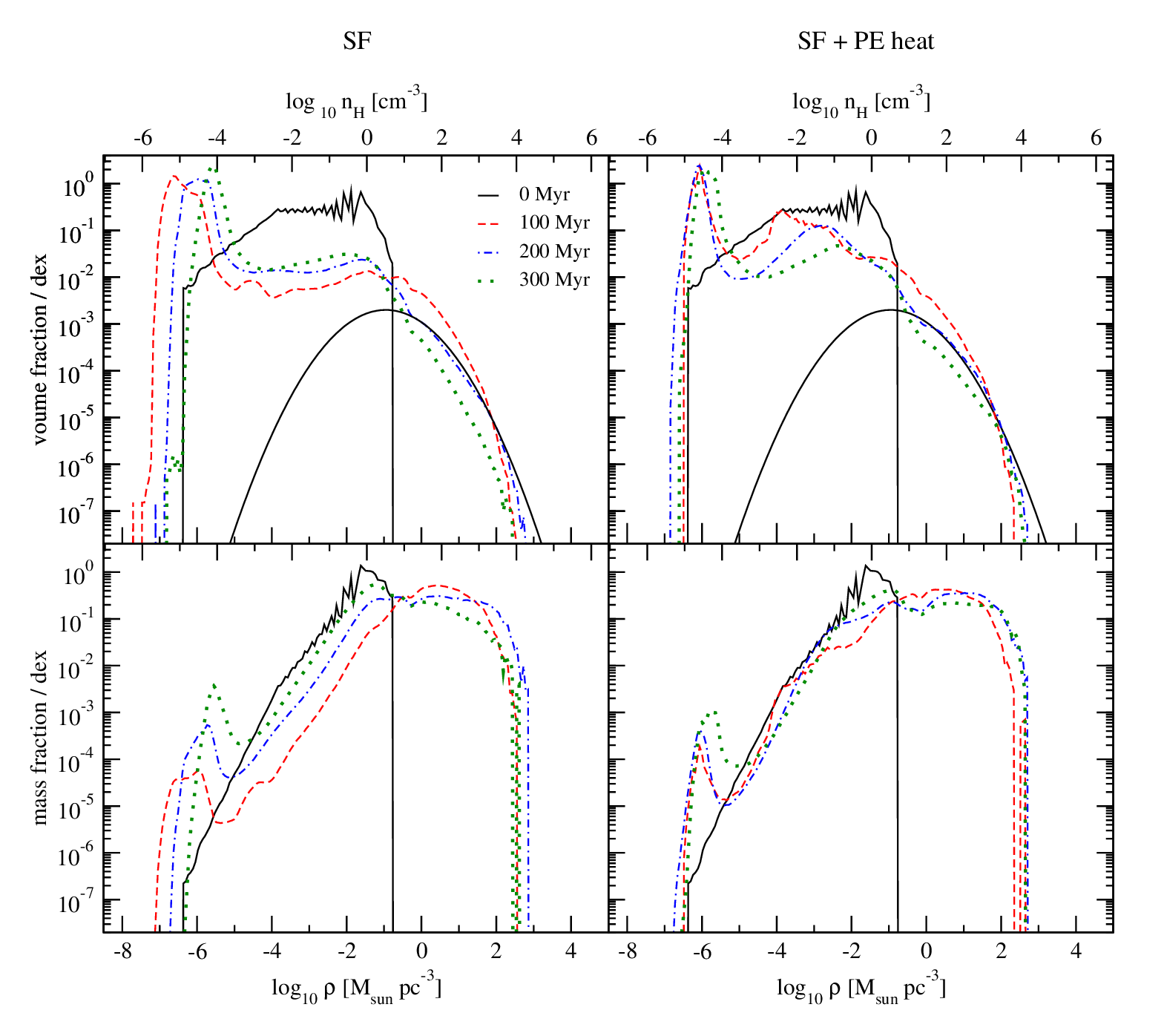}
\caption{Probability distribution function (PDF) for the disks over the
  radii $2.5 < r < 8.5$\,kpc and height $-1~{\rm kpc}<z<1~{\rm
    kpc}$. Left-hand plots are for the disk with only star formation
  (disk SFOnly)
  right-hand plots are for the disk where diffuse heating is
  included (disk SF+PEheat). The top panel shows the evolution of the
  volume-weighted PDF where the solid-line curve shows a log normal
  fit to the high density PDF tail. The bottom panel is the
  mass-weighted PDF. There is relatively little evolution in the PDF shape over the
  course of the simulation, although the depletion of gas in the SFOnly
  disk is evident in the gas fraction shift to lower densities. This
  is not as evident in SF+PEheat and more mass exists in
  the mid-density ($10^{-4} < n_H <1$\,cm$^{-3}$) ISM. 
\label{fig:diskpdf}}
\end{center} 
\end{figure*}

\begin{figure*} 
\begin{center} 
\includegraphics[width=\textwidth]{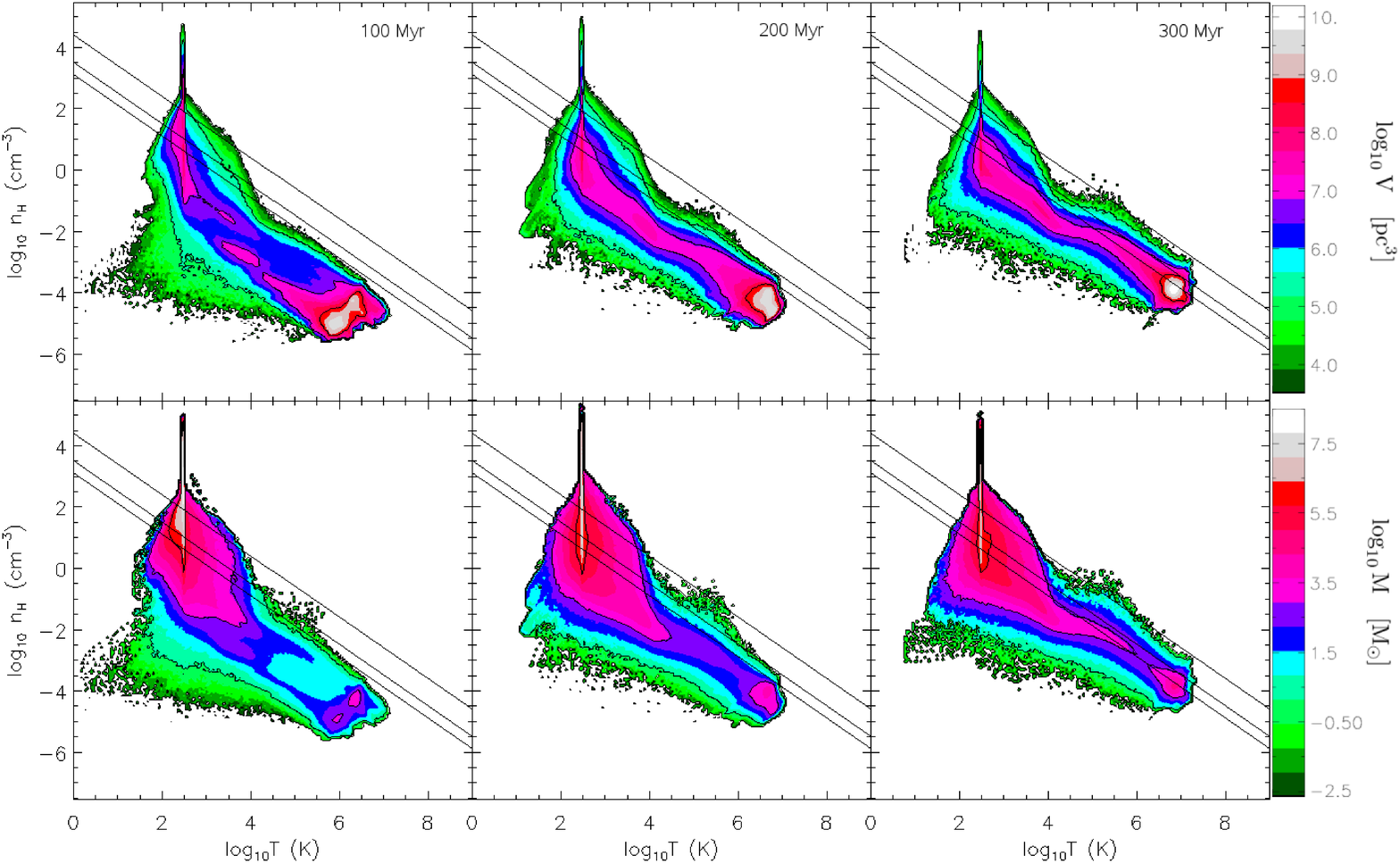}
\includegraphics[width=\textwidth]{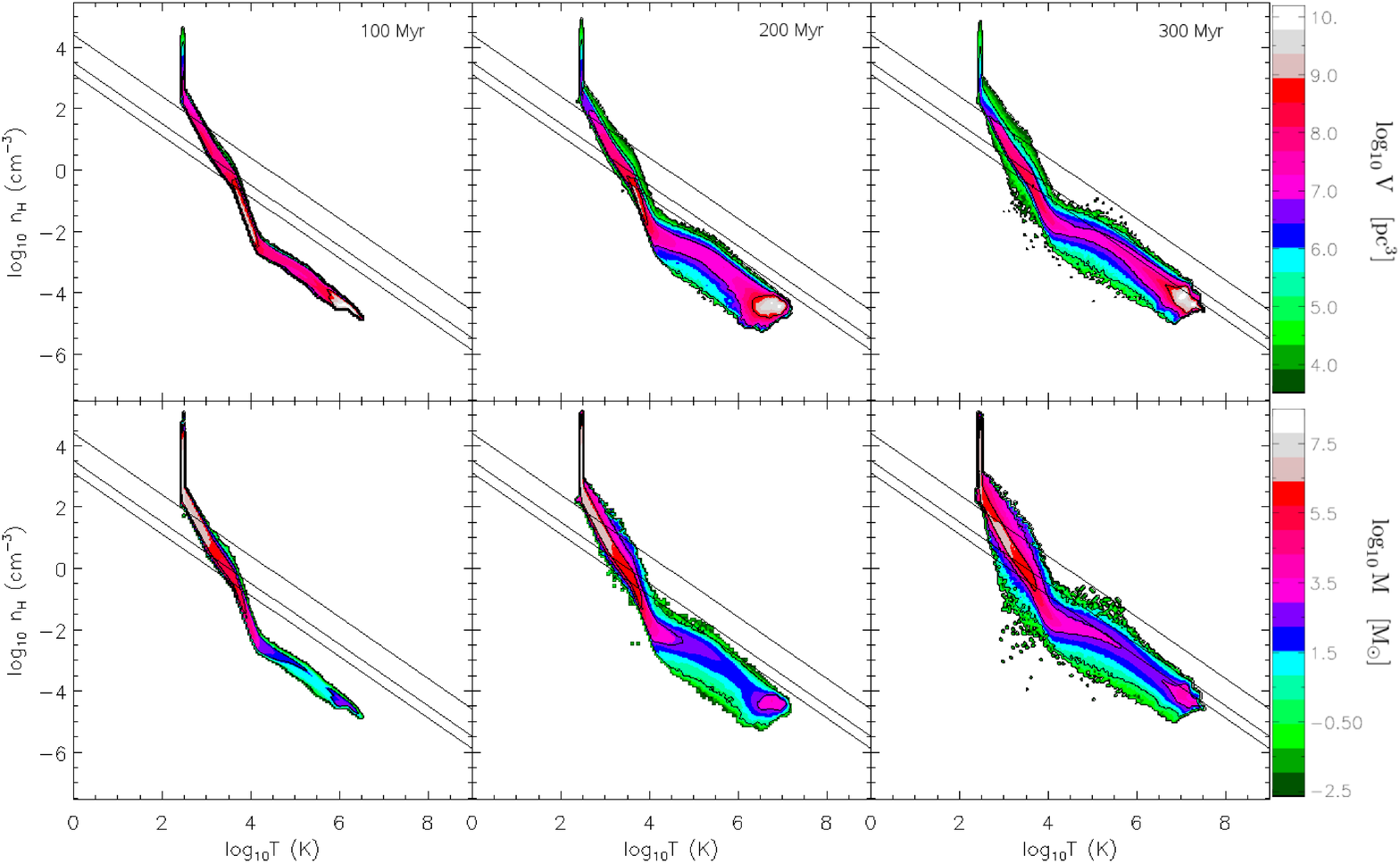}
\caption{Density vs temperature contour plots for disk SFOnly (upper
  panel of six images) and disk SF+PEheat (lower panel of six). In each panel, the top row
  shows the distribution of gas volume while the bottom row shows gas
  mass for the region $2.5~{\rm kpc}<r<8.5~{\rm kpc}$ and $-1~{\rm
    kpc}<z<1~{\rm kpc}$. Solid lines show the total pressure in the
  Milky Way, $P_{\rm tot}/k_b= 2.8 \times 10^4$\,K cm$^{-3}$ (top),
  the thermal pressure, $P_{\rm th}/k_b = 0.36 \times 10^4$ K
  cm$^{-3}$ (middle) and the thermal pressure excluding the hot gas
  component, $P_{\rm th, no hot}/k_b = 0.14 \times 10^4$ K cm$^{-3}$
  (bottom) \citep{Boulares1990}. We see a continuous range of
  densities and pressures in both disks that is in pressure
  equilibrium above temperatures of $10^4$\,K and over-pressurized in
  the self-gravitating clouds. Diffuse heating greatly reduces the
  range of values due to the lower velocity dispersion and denser warm ISM.
\label{fig:contours}}
\end{center} 
\end{figure*}

A one dimensional analysis of the structure of the ISM in the main
region of the disk can be seen in the probability distribution
functions (PDFs) plotted in Figure~\ref{fig:diskpdf} for the
simulations at the same times shown in Figure~\ref{fig:diskevol}. The
top two graphs show the gas volume fraction as a function of density
for the disk SFOnly (left) and for disk SF+PEheat (right). The bottom graphs
show the mass fraction over the same density range. 

Without diffuse heating, the mass fraction of gas in disk SFOnly above
our cloud definition limit of $n_{\rm H,c} > 100$\,cm$^{-3}$ is 0.43, 0.4 and 0.16
for simulation times $t = 100, 200$ and 300\,Myr respectively. For disk
SF+PEheat, the same output times have cloud mass fractions of 0.36,
0.46, 0.33. The rise in dense gas from 100 to 200\,Myr when diffuse
heating is present is evidence for the disk still fragmenting over
this period. 

Little evolution in the shape of the PDFs is seen over the course of the simulations once the initial
conditions have produced a fragmented disk. In contrast to disk NoSF in TT09, we do not see a rise in high
density gas over time, since stars now form in gas above
$n_{\rm H,c} >
100$\,cm$^{-3}$ and deplete the gas reservoir above this
threshold. This will have an impact on the maximum cloud mass as will
be seen in section~\ref{sec:cloudproperties}.

In agreement with other simulations of galactic disks
\citep[e.g.][]{Robertson2008, Tasker2008, Wada2007}, both our disks
can be fitted with a log-normal tail to their volume-weighted
PDFs. In disk NoSF in TT09, the fit to the 200\,Myr and 300\,Myr lines
extended up to densities of $n_{\rm H} < 10^5$\,cm$^{-3}$, but in these cases star
formation removes gas to steepen the profile at the high density
tip, making this fit not as good. This log normal fit from disk NoSF
is shown in Figure~\ref{fig:diskpdf} and has the form,

\begin{equation}
{\rm PDF} =
\frac{1}{\sigma_{\rm PDF}\sqrt{2\pi}}e^{-\left(\ln x -\bar{\ln x}\right)^2/2\sigma_{\rm PDF}^2},
\end{equation}

where $x = \rho/\bar{\rho}$ and $\sigma_{PDF} = 2.0$. It remains a
good fit to both disks at $t = 200$\,Myr, but by $t = 300$\,Myr, star
formation has significantly eroded the dense gas. 

The one-dimensional Mach number, ${\cal M}$, for the star-forming high density tail
of the PDF can be estimated from the azimuthally averages profiles in
Figure~\ref{fig:diskevol}. Before star formation significantly
depletes the disk gas, the velocity dispersion, $\sigma_g$, is
dominated by the in-place velocity motions, $\sigma_{\rm nt}$, which
are of order $15$\,kms$^{-1}$. In this high density region, the
temperature of the gas is close to our cooling floor at 300\,K (a fact that will
also be seen in Figure~\ref{fig:contours}), giving a $c_s =
1.8$\,kms$^{-1}$. The Mach number is given by the ratio of these two
values, ${\cal M} = \sigma_{\rm nt}/c_s \approx 8.3$.

From this and the value of $\sigma_{PDF}$, the nature of the
turbulence production in the star-forming gas can be deduced. The two
numbers are related via $\sigma^2_{\rm PDF} = \ln\left[1+b^2{\cal
    M}^2\right]$, where $b$ is found to vary between $b\sim 1/3$ for
solenoidal (divergence-free) turbulent modes and $b\sim 1$ for compressive
(curl-free) modes \citep{Federrath2008}. For ${\cal M}
= 8.3$ and $\sigma_{PDF} = 2.0$, we find $b =0.88$, suggesting that
compressive forcing is the dominant turbulent mode. This is consistent
with the compressive nature of cloud collisions which are likely to be
a driving force for the turbulence. 

Removing gas from the high density tail of the PDF via star formation
causes an increase in the fraction of gas at densities below $n_H
<1$\,cm$^{-3}$. This effect is more pronounced in disk SFOnly, since
the fraction of cloud gas has dropped by almost
a factor of 3 between 100 and 300\,Myr, compared with a maximum change
of just 1.4 in disk SF+PEheat.

Despite its temporal continuity, the fraction of gas between $10^{-4} < n_H
<1$\,cm$^{-3}$ in disk SF+PEheat is higher than for disk
SFOnly, remaining close to the initial condition value. This gas
is the warm ISM in which the clouds are embedded and, as seen in
Figure~\ref{fig:diskimages}, has a denser filamentary structure due to
a smaller level of fragmentation.

The peak at low densities of $n_H < 10^{-4}$\,cm$^{-3}$ corresponds to
gas in the 1\,kpc region above and below the disk plane. In the disk
SFOnly, we see the density of this region increase
over time as the high velocity dispersion seen in
Figure~\ref{fig:diskevol} causes the disk scale height to
increase. With diffuse heating, the disk settles to an initially
thicker profile with a lower velocity dispersion in its less
fragmented ISM, undergoing less evolution.

A two-dimensional representation of the evolution of the ISM is shown
as density vs temperature contour plots in
Figure~\ref{fig:contours}. The upper two rows of plots show the gas volume
distribution (top) and gas mass distribution (bottom) at times $t = 100,
200$ and 300\,Myr for disk SFOnly while the lower
six plots show the same distributions for disk SF+PEheat. Diagonal
lines mark constant pressure with the values
being representative of the Milky Way \citep{Boulares1990}: The
upper-most line is the estimated value for the total Milky Way
pressure, $P_{\rm tot}/k_b= 2.8 \times
10^4$\,K cm$^{-3}$. The lower two lines show the thermal pressure,
$P_{\rm th}/k_b = 0.36 \times 10^4$ K cm$^{-3}$, and the thermal
pressure excluding the hot gas component, $P_{\rm th, no hot}/k_b = 0.14 \times 10^4$ K cm$^{-3}$.

The distributions show that gas warmer than $10^4$\,K, the typical
temperature for the warm interstellar medium \citep{McKee1977}, lies
largely in pressure equilibrium, although with a wide range of values
that run over a continuous distribution of densities and
temperatures. Without diffuse heating, gas in disk SFOnly that is cooler than this remains in
pressure equilibrium until it approaches the floor in our radiative
cooling curve at $300$\,K. This appears as a sharp spike in high mass
and low volume density in all plots. The gas in this region is in the
star forming clouds whose self-gravity causes them to be over
pressurized with respect to the surrounding ISM. There is a
perceptible decrease of gas in this spike over time as the cloud gas
is converted into stars. Temperatures lower than the cooling
curve minimum are achieved via adiabatic expansion. 

With the addition of diffuse heating in disk SF+PEheat, cooler gas is warmed causing its
pressure to increase. This results in a significantly thinner profile
and gas below $T < 10^4$\,K being out of pressure equilibrium. The
smaller range of temperatures and densities in the warm ISM is
indicative of the coherent structure of filaments we saw in
Figure~\ref{fig:diskimages} and the lower velocity dispersion in
Figure~\ref{fig:diskevol}.  

Overtime, both disk profiles broaden as cloud mergers and tidal
encounters disrupt the structure of the dense gas. 

For the gas in pressure equilibrium at $T > 10^4$\,K, its value
is lower than that of the Milky Way in both the disks during the
majority of the simulation. This is not surprising, since gas at this
temperature will be strongly affected by sources of energetic feedback
which we do not consider. We do see a rise in the pressure over time,
resulting in a factor of 10 increase between $t = 100$ to
300\,Myr. Comparing with the results in disk NoSF, where no such increase
was seen, we conclude this is the result of the localized potential created by the star particles.

\section{The Properties of the Clouds}
\label{sec:cloudproperties}

\begin{figure*}[!t] 
\begin{center} 
\includegraphics[angle=270,width=\textwidth]{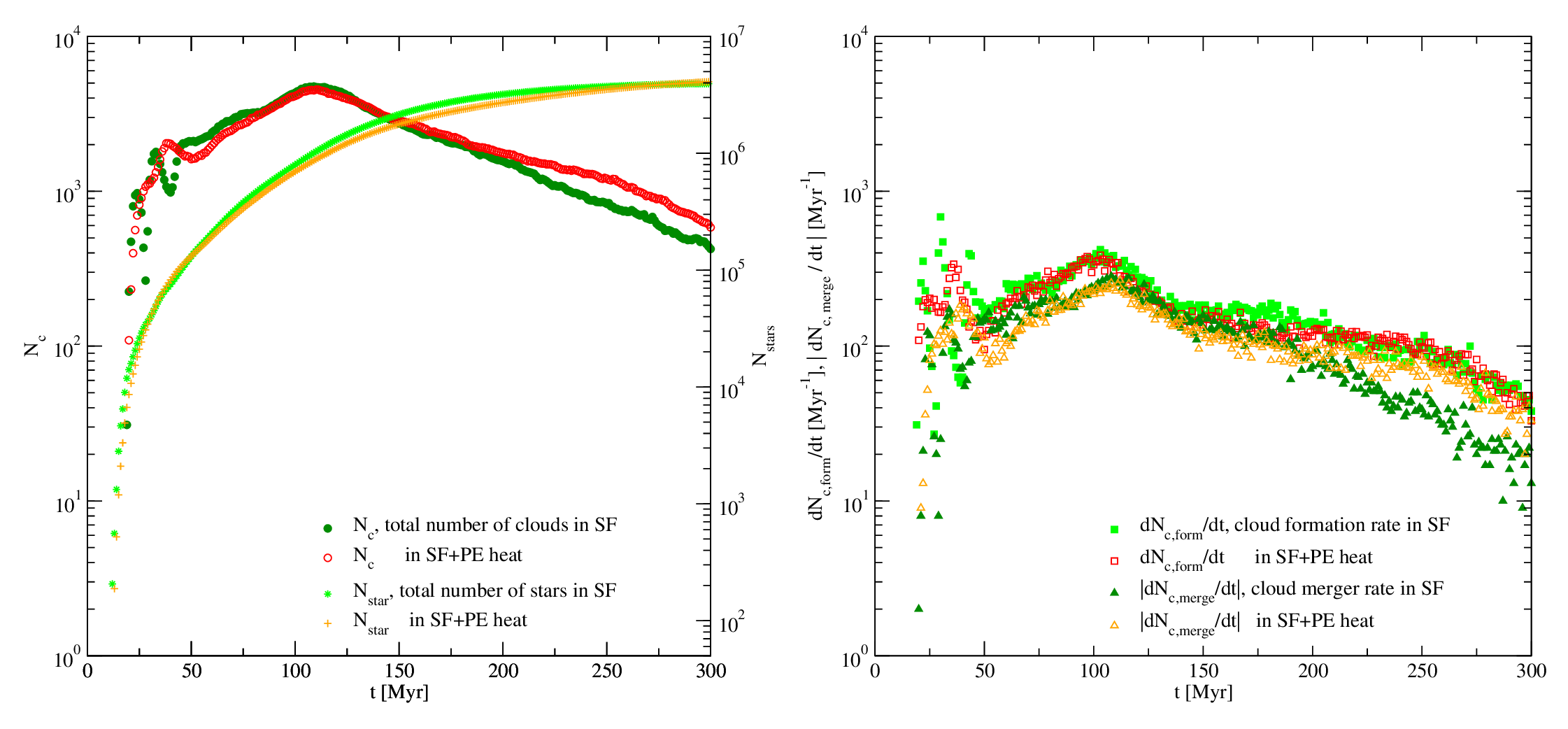}
\caption{Left: Evolution of the total number of GMCs, $N_c$, (on left
  axis) and total number of star particles, $N_{\rm stars}$, (right axis) for both
  disks. Right: the GMC formation rate, $dN_{c,form}/dt$, and the GMC merger rate (equivalently, the rate of
  cloud destruction via mergers, since these are virtually all between
two clouds), $|dN_{\rm c,merge}/dt|$ for both disks. The disk fragments over the first 100\,Myr where the number
of clouds begins to decrease due to mergers between clouds and gas depletion
from star formation. The latter process is slower in the gas with
diffuse heating, which suppresses the fragmentation of the disk and
star formation in the clouds.)
\label{fig:formation_history}}
\end{center} 
\end{figure*}

\begin{figure*} 
\begin{center} 
\includegraphics[width=8cm]{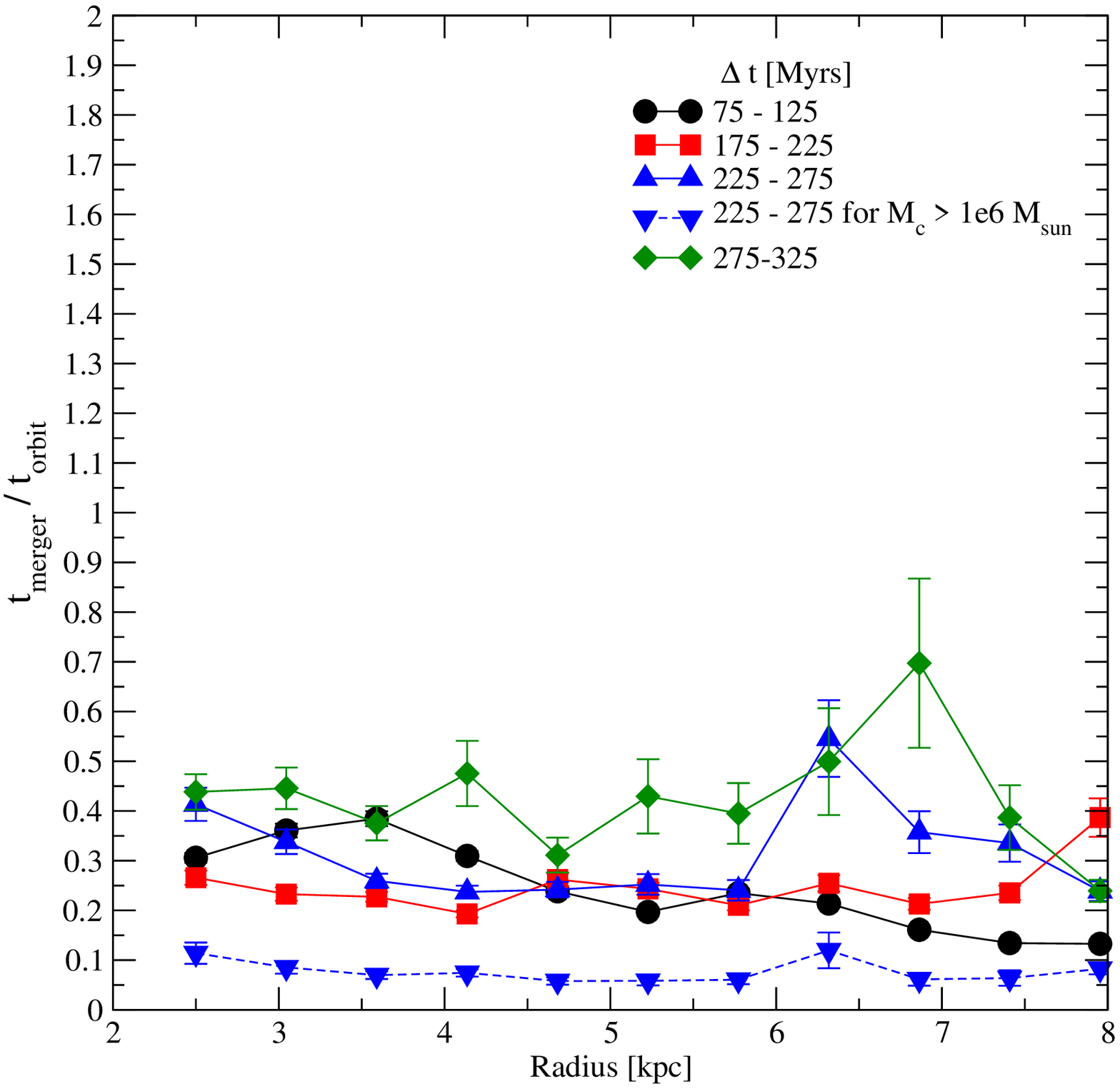}
\includegraphics[width=8cm]{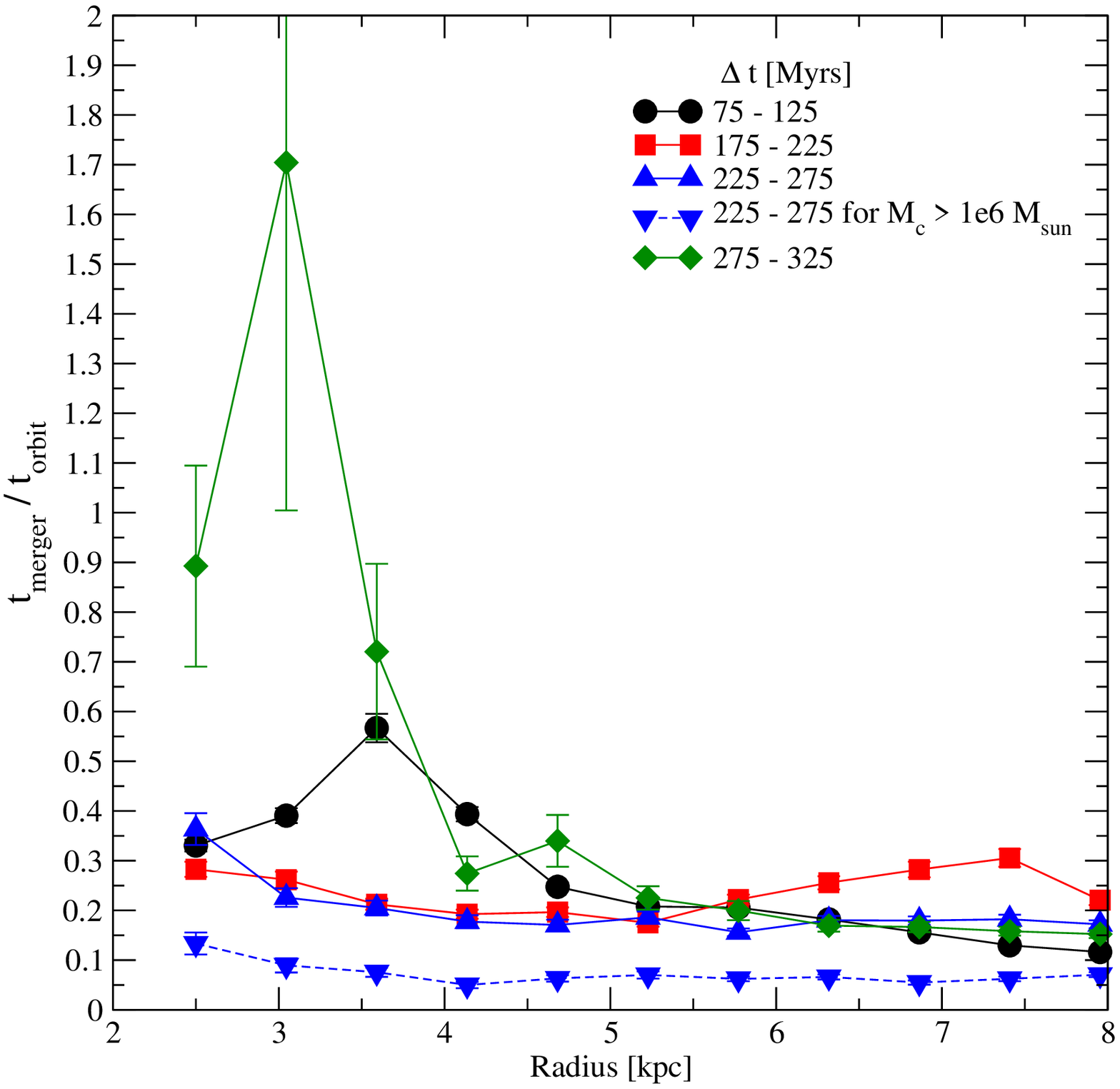}
\caption{Cloud merger timescales, averaging over 50\,Myr intervals of
  simulation time, for disks SFOnly (left) and SF+PEheat (right). Only
  clouds born after 140\,Myr, the initial fragmentation of the disk,
  are included in the analysis. The average time for a merger is low
  at $\sim 0.25$ of an orbital period
and largely independent of galactocentric radius. The dashed line
shows the merger times of clouds with $M_c > 10^6$\,M$_\odot$ (i.e., the average time for a cloud of that
size to undergo a merger with a cloud of any mass), evaluated over the
interval t = 225-275 Myr, which is lower by a factor of 2.
\label{fig:mergers}}
\end{center} 
\end{figure*}

While the galaxy's global structure concerns the entire ISM, its star
formation occurs almost solely in the densest, coldest component of
that gas. This cold phase is organized into the extended structures
of the GMCs. As the disk cools and fragments, regions of gas exceed
our threshold of 100\,cm$^{-3}$ and we recognize them as the GMCs. The
properties of these entities dictate the environment in which stars
will form and their birth, evolution and death will determine the star
formation rate in the galaxy. 

In this section, we focus on the properties of the identified GMCs,
examining the the evolution of the clouds both as a function of
simulation time, $t$, and as their age, or cloud time, $t'$. 

The evolution of the cloud population itself is shown
Figure~\ref{fig:formation_history} for both disks. The left-hand plot
shows the total number of clouds and stars in the disk, while the
right-hand plot shows the formation and merger rate of the clouds. The number
of clouds initially increases as the disk gravitationally fragments,
peaking at a rate of around $400$\,Myr$^{-1}$ at 100\,Myr in both
disks. After that time, the disk is fully fragmented and both the
cloud population and formation rate drop, accompanied by an increase
in the number of star particles, as gas is consumed during star formation and
mergers further erode the cloud population. There is a delay in the
peak of the cloud formation rate and cloud merger rate that is
equivalent to the mean merger rate between clouds. By 300\,Myr,
the cloud population has dropped by a factor of 10 without diffuse
heating and a factor of 8 where diffuse heating was included. When
diffuse heating is present, the build up of the stellar population is
slightly more gradual, indicative of a slower star formation rate. By 300\,Myr,
the total number of star particles in both disks is approximately the same at
$\sim 4$\,million, but at 200\,Myr, it differs by $\sim 20$\%. This
accounts for the higher number of clouds in disk SF+PEheat at
300\,Myr; the slower star formation rate allows the cloud time to
accrete material as it is consumed, allowing a greater number to
survive despite the ultimate star particle number being the same as in
disk SFOnly.

Mergers between clouds happen throughout the course of the simulation,
their rate related to the number of clouds in the disk. How mergers
are recorded by the tracking algorithm is discussed in TT09, but in
short, a merger is said to have happened when a single cloud is at the
predicted position for two other clouds after 1\,Myr of evolution,
within a margin of twice the average radius of the potential merger
product. 

In the disk SFOnly, the number of mergers drops
more steeply in the second half of the simulation compared to
SF+PEheat. This corresponds to a steeper fall
off in the number of clouds as the mass of gas above the cloud
threshold density drops heavily as was seen in Figure~\ref{fig:diskpdf}. As we
will see in Section~\ref{sec:stars}, gas depletion in this simulation
also causes the star formation rate to drop considerably over this
period compared to the diffusely heated disk.

The frequency of the mergers in both disks are shown in
Figure~\ref{fig:mergers} as a function of the galactocentric
radius. The merger time is computed as a fraction of the orbital time
and averaged over 50\,Myr. Merger rates over four time
intervals during the simulation (75-125\,Myr, 175-225\,Myr,
225-275\,Myr, 275-325\,Myr) are plotted, where a merger is considered
to be between two clouds. Cloud mergers involving more than two
objects are negligibly rare.  

There is no trend between merger rate and galactocentric radius for
the majority of the simulation. At late times, disk SF+PEheat
undergoes less mergers in its inner 4\,kpc, corresponding to
the drop in gas density (and so cloud number density) in that region,
coupled with the lower velocity dispersion, as seen in
Figure~\ref{fig:diskevol}. In disk SFOnly, there is a spatially uniform increase at latest
simulation time in the time between mergers, due to the
reduced number of clouds seen in Figure~\ref{fig:formation_history}.

Within the first three time frames, the average number of mergers is
approximately constant in both simulations, with a value of $\sim 0.25$ of the
orbital time, close to that seen in TT09 for disk NoSF. For our largest clouds, $M >
10^6$\,M$_\odot$, this rate is lower by a factor of 2. This is in agreement with
the calculations of \citet{Tan2000}, who estimated $t_{\rm
  merger}/t_{\rm orbit} = 0.2$ in an analytical model. Since the disk
circular velocity is $v_c = 200$\,kms$^{-1}$, the average time between
interactions is $\sim 30$\,Myr at a radii of 4\,kpc. This is slightly higher
than for the clouds in disk NoSF where an average time of 25\,Myr was found, due to the
star formation depleting the number of clouds. 

The frequency of the cloud collisions is indicative that they play a
major role in shaping the properties of the GMCs. What is less clear
is what effect they have on the star formation rate as will be
discussed in Section~\ref{sec:stars}.

\subsection{GMC Properties with Simulation Time}
\label{sec:simtime}

\subsubsection{Simulation Time: Cloud Property Distributions}
\label{sec:simtime_distributions}

\begin{figure*} 
\begin{center} 
\includegraphics[width=15cm]{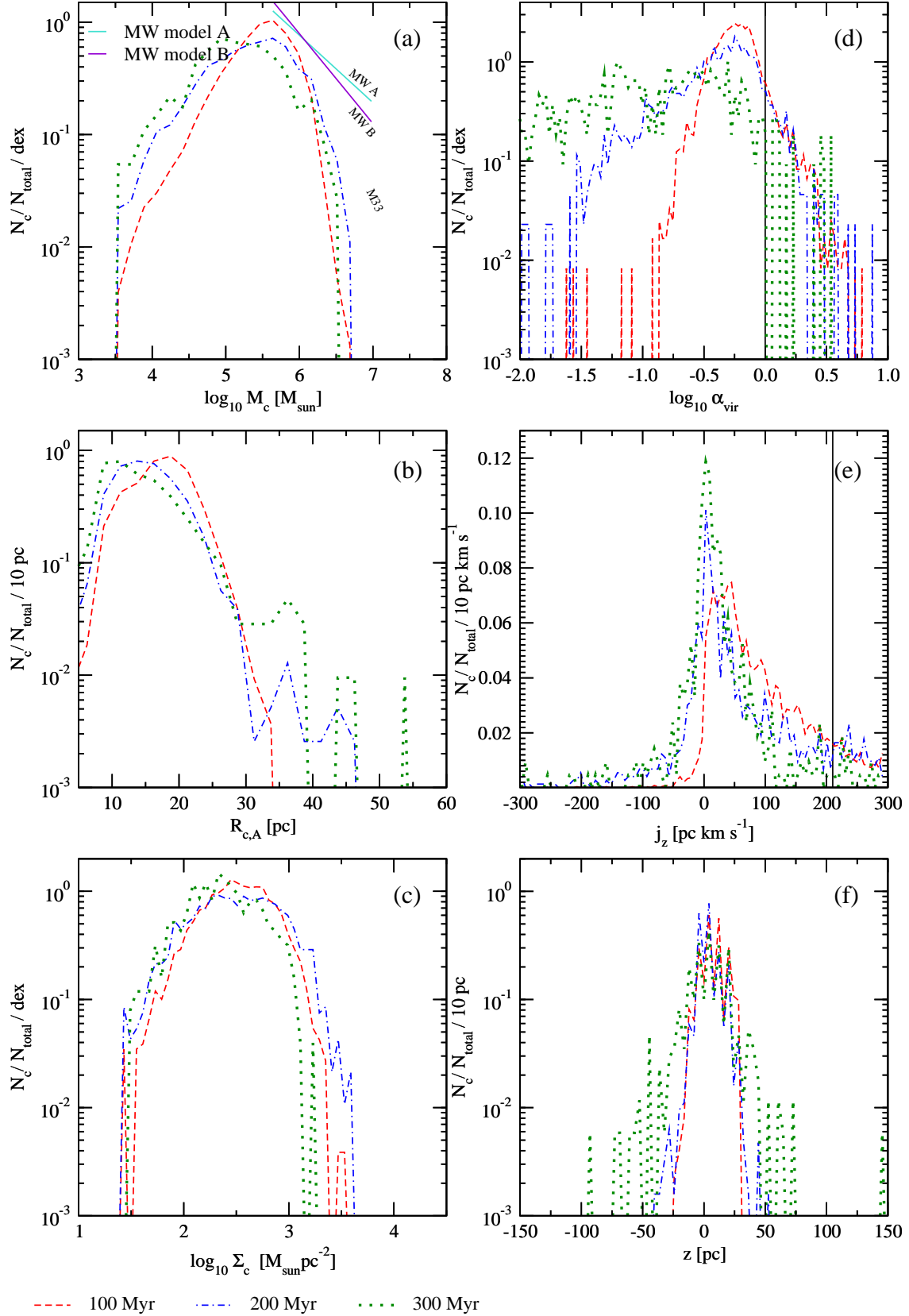}
\caption{Normalized distributions of GMC properties at 100, 200 and 300\,Myrs
  after the start of the simulation for disk SFOnly. Top left plot
  shows: (a) cloud mass distribution overlaid with fits to
  observational data from the Milky Way (blue and violet solid lines;
  \citet{Williams1997}). Middle left, (b): the average
  radii of clouds, $R_{c,A} \equiv (A_c/\pi)^{1/2}$ where $A_c$ is the
  projected area of the clouds in the Y-Z plane. Bottom left, (c): mass surface
  density of the clouds, $\Sigma_{c,A}$. Top right, (d): virial parameter,
  $\alpha_{\rm vir}$ with the vertical line indicating where
  $\alpha_{\rm vir} = 1$, the limit for gravitational binding. Middle right, (e):
  vertical component of the specific angular momentum vector,
  $j_z$. The vertical line indicates a value of $j_z$ of a spherical
  ($\simeq 110$\,pc radius) region of the initial conditions at
  galactocentric radius $r = 4$\,kpc containing
  $10^6$\,M$_\odot$. Bottom right, (f):
  vertical height distribution, $z$, of the clouds.}
\label{fig:simtime_clouds_sf5}
\end{center} 
\end{figure*}

\begin{figure*} 
\begin{center} 
\includegraphics[width=15cm]{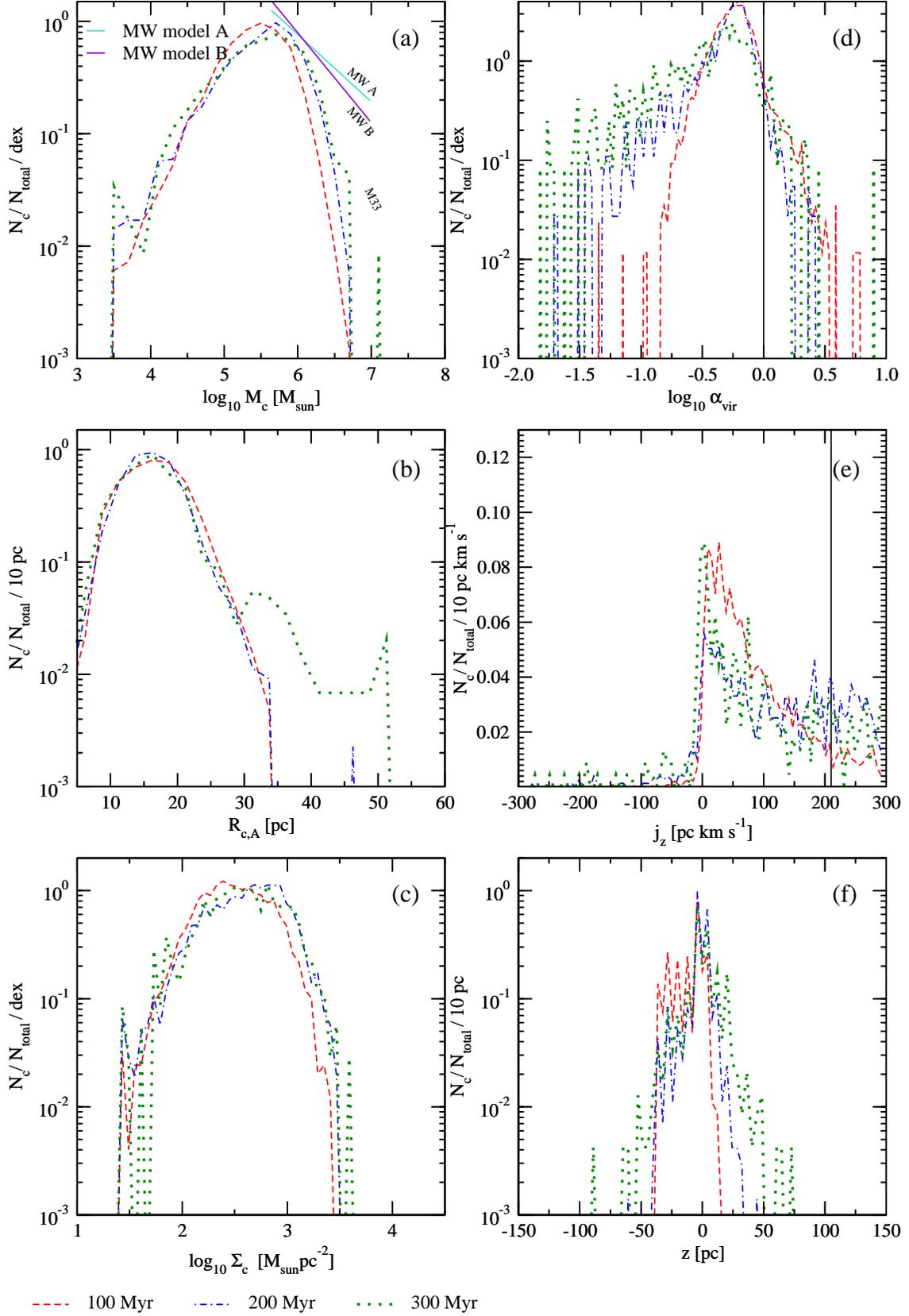}
\caption{Normalized distributions of GMC properties at 100, 200 and 300\,Myrs
  after the start of the simulation for disk SF+PEheat. Plots are the same as those
  described in Figure~\ref{fig:simtime_clouds_sf5}. Top left: (a) cloud
  mass, $M_c$. Middle left: (b) cloud radius, $R_{C,A}$. Bottom left:
  (c) mass surface density, $\Sigma_c$. Top right (d) virial
  parameter, $\alpha_{\rm vir}$. Middle right (e) vertical component
  of the specific angular momentum, $j_z$. Bottom right: (f) cloud
  center-of-mass vertical positions, $z$.}
\label{fig:simtime_clouds_peheat}
\end{center} 
\end{figure*}

The properties of the individual clouds are shown in
Figure~\ref{fig:simtime_clouds_sf5} for disk SFOnly and
Figure~\ref{fig:simtime_clouds_peheat} for the clouds
for disk SF+PEheat. The distributions are shown at three
different times during the simulation; 100 (red dashed lines), 200
(blue dot-dashed lines) and 300\,Myr (green dotted lines).

The top left plot, Figure~\ref{fig:simtime_clouds_sf5}\,(a) and
Figure~\ref{fig:simtime_clouds_peheat}\,(a), show the
profiles for the cloud gas mass. In difference to when no star
formation occurred in disk NoSF and in agreement with the PDF plots in
Figure~\ref{fig:diskpdf}, we do not see an increase in the cloud mass
over the course of the simulation. Instead, star formation prevents
the formation of a high mass tail and caps the maximum cloud mass at $M <
10^{6.7}$\,M$_\odot$. When diffuse heating is included in
Figure~\ref{fig:simtime_clouds_peheat}\,(a), we see one outlier beyond
this truncation at $M = 10^{7.1}$\,M$_\odot$ which could be due to
a recent merger event. Otherwise, the addition of heating does not
change the mass cut-off. 

The peak of the mass distribution appears at  $10^{5.6}$\,M$_\odot$ at 
100\,Myr after the start of the simulation in disk SFOnly and slightly
lower, $10^{5.5}$\,M$_\odot$, in disk SF+PEheat. This drop in peak
mass is due to the heating term
supporting the cloud against its own gravity, slowly the collapse to
its maximum density. This was noted in the disk PDFs in
Figure~\ref{fig:diskpdf} and will be seen again below in the
distribution of cloud surface densities.

Without diffuse heating, the conversion of gas into stars causes the
peak to migrate to lower masses over time, moving to
$10^{4.9}$\,M$_\odot$ by 300\,Myr. At this stage, 75\,\% of the clouds
have a stellar mass fraction greater than 90\,\%. This high stellar content
is due to the lack of feedback that would act to prevent gas
continually collapsing to form stars within the cloud. With diffuse
heating, the evolution of the mass profile is reduced, with no
decrease in the peak position over time and a smaller evolution in the
low mass end of the spectrum. This is in
keeping with Figure~\ref{fig:diskpdf} where we saw little change in
the heated disk's ISM over time. Again, the added thermal support is
the cause here, decreasing the cloud density and so increasing its
dynamical time, causing a slower rate of conversion from gas into
stars through equation~\ref{eq:starmass}.

Observations of the Milky Way by \citet{Williams1997} find a
truncation in the cloud molecular mass at $M_{\rm
 u} \sim 6\times 10^6$\,M$_\odot$. Since we follow all the neutral
gas, our results also include an atomic envelope whose mass is
estimated by \citet{Blitz1990} to be of order the molecular
component. This would take the observed truncation limit up to $M <
1.2 \times 10^{7}$\,M$_\odot$. Recent observations by
\citet{Fukui2009} of the LMC, suggest smaller envelopes with 10\% of the
mass, providing a limit of $M_{\rm u} \sim 6.6\times
10^6$\,M$_\odot$. In both cases, our maximum mass falls below or close
to the observed truncation.

This simulated mass spectrum can be compared to observations of GMCs
in the Milky Way. These observations \citep[and indeed, those in other
galaxies such as M33 as found by][]{Rosolowsky2003} fit a power law
of the form $\frac{dN_c}{d\ln M_c} \propto M_c^{\alpha_c}$. The two lines on
Figure~\ref{fig:simtime_clouds_sf5}\,(a) and
Figure~\ref{fig:simtime_clouds_peheat}\,(a) show this relation for
observations of $\alpha_c$ in the Milky Way made by
\citet{Williams1997}. Their measured value for $\alpha_c$ depends
on if the observed sample is assumed to be equally under-sampled at all masses
($\alpha_c = 0.6$, MWA) or if lower mass clouds are more under-sampled than
high mass ($\alpha_c = 0.8$, MWB).

The distributions most appropriate to compare with these observations
are the lines at 200 and 300\,Myr, after the disk has settled and
cloud interactions have become the dominant force in the ISM.  Within
the mass range $\sim (0.5 - 10)\times 10^6$\,M$_\odot$, where our
clouds are most resolved, we find that disk SFOnly approaches the
Milky Way GMC population. With diffuse heating,
the evolution of the mass profile is reduced, and the distribution is
steeper at later times. 

\citet{Williams1997} estimate that, within the inner Milky Way,
there are $\sim 1000$ clouds with $M_c > 10^5$\,M$_\odot$ and 100-200 clouds with
$M_c > 10^6$\,M$_\odot$. At 200\,Myr, we find for disk SFOnly 1122 clouds with $M_c > 10^5$\,M$_\odot$ and 179 clouds with
$M_c > 10^6$\,M$_\odot$, in good agreement with this result. The cloud
population in disk SF+PEheat is slightly larger with 1491 clouds with $M_c >
10^5$\,M$_\odot$ and 238 with  $M_c > 10^6$\,M$_\odot$, reflecting the
higher fraction of cloud gas at this time, but is still of the same
order as the observations. Moreover, the mass fraction of clouds at 200\,Myr is 0.4 for the disk
without diffuse heating and 0.46 when it
is included (as seen in section~\ref{sec:ism}), coinciding
with the estimated value of \citet{Wolfire2003} for the fraction of
molecular to atomic gas inside the solar radius. 

Figure~\ref{fig:simtime_clouds_sf5}\,(b) and
Figure~\ref{fig:simtime_clouds_peheat}\,(b) show the average radius of
the clouds defined as $R_{c,A} \equiv (A_c/\pi)^{1/2}$, where $A_c$ is
the projected area of the cloud in the $y-z$ plane (that is, as it would
be measured by an observer embedded in the plane of the galaxy). Cloud
radii are typically around 15\,pc in both disks, in good agreement with
observations of GMCs in the Milky Way and M33 \citep{Lada2010,
  Rosolowsky2003}. This is slightly smaller than the
non-star forming clouds in disk NoSF (typical
radius $\sim 20$\,pc), but the inclusion of star formation has only
significantly impacted the tail of the distribution which, as with the
mass, is now prevented from continual growth. 

As with the mass profile in Figure~\ref{fig:simtime_clouds_sf5}\,(a),
there is a decrease in the peak cloud radius over time for disk
SFOnly, as gas is converted into stars. This is not
reflected in disk SF+PEheat as its cloud mass fraction
changes slowly under the influence of a lower SFR. At later times,
there are a small number of extended structures which are likely to be
from older clouds that have undergone multiple merger events. 

Figure~\ref{fig:simtime_clouds_sf5}\,(c) and
Figure~\ref{fig:simtime_clouds_peheat}\,(c) show the surface density of
the clouds, $\Sigma_c \equiv M_c/A_c$. The distribution is the most
robust property plotted, showing minimum evolution over
time. The peak value sits $\sim 300$\,M$_\odot$pc$^{-2}$, in agreement
with disk NoSF, although at 100\,Myr, disk SFOnly has a
lower higher peak surface density than disk SF+PEheat, as the
heating slows down the fragmentation to the highest densities. 

There is a minor amount of evolution over time. Without diffuse
heating, the surface densities drop as the clouds becomes more stellar
dominated. With the diffuse heating, the shift is in the opposite
direction as the disk self-gravity slowly overwhelms the support from
heating and amount of cloud gas rises between 100\,Myr and
200\,Myr. Compared to the changes in mass and radius profile, however,
these shifts are small.

The independence of the surface density profile to physics and time agrees with \citet{Larson1981}
and \citet{Solomon1987}, who found that
GMCs in the Milky Way had a constant surface density. Their
measurements give a value for $\Sigma_c$ lower than what we find, with
$\Sigma_c = 200$\,M$_\odot$pc$^{-2}$, but this does not take into
account the atomic envelope, which can be a factor of 100 lower in
density \citep{Fukui2009}. 

The degree of gravitational binding in the clouds can be estimated by
the alpha virial parameter, $\alpha_{rm vir}$, plotted in
Figure~\ref{fig:simtime_clouds_sf5}\,(d) and
Figure~\ref{fig:simtime_clouds_peheat}\,(d). This quantity is defined as
$\alpha_{vir} \equiv 5\sigma^2_cR_{c,A}/(GM_t)$, where $\sigma_c$ is
the mass averaged velocity dispersion of the cloud, $\sigma_c \equiv
(c_s^2 + \sigma_{nt,c}^2)^{1/2}$, with $\sigma_{nt,c}$ the
one-dimensional rms velocity dispersion about the cloud's
center-of-mass velocity \citep{Bertoldi1992}. In contrast to TT09,
$M_t$ is now the sum of the total gas and stellar mass in the
cloud. To calculate this, star particles are associated with the cloud if they
reside within its boundaries or within a distance of twice its average
radius. A value of $\alpha_{\rm vir} = 1.0$ states that a spherical,
uniform cloud with negligible surface pressure and magnetic fields is
virialized. This translates to a cloud for which $\alpha_{\rm vir} <
1.0$ being gravitationally bound and one for which $\alpha_{\rm vir} >
1.0$ likely to be unbound. Observationally, clouds are seem to be on
the borderline of the two states, with $\alpha_{vir} \sim 1$
\citep{McKee2007}. 

For both our disks, the profile peaks at $\log(\alpha_{vir}) \sim -0.25$
throughout the simulation, implying our clouds are
weakly bound. There is an increase in the fraction of more highly
bound objects over time as the clouds become more stellar dominated,
resulting in mergers increasing the cloud's total mass more than its
gas fraction (which would increase its radius or surface
density). This effect is strongest in disk SFOnly, with the
distribution at 300\,Myr showing an almost flat
profile below $\alpha_{\rm vir} < 0.5$. As previously mentioned, 75\%
of the clouds at this time are now over 90\% stellar in mass, the
relatively small quantity of gas making it unlikely that our expression for $\alpha_{vir}$ can
still apply. With diffuse heating this fraction is only $\sim 46\%$
and 300\,Myr, and $\alpha_{\rm vir}$ shows a smaller level of evolution.

The specific angular momentum in the vertical, $z$, component is shown
in Figure~\ref{fig:simtime_clouds_sf5}\,(e) and
Figure~\ref{fig:simtime_clouds_peheat}\,(e).  The vertical line marks
the specific angular momentum that a sphere of $10^6$\,M$_\odot$ and radius
$r = 4$\,kpc has in the initial conditions. The clouds have an angular
momentum much smaller than this, since they are much more compact
objects and therefore less susceptible to the disk sheer.

Without diffuse heating, the angular momentum profile in disk SFOnly is very similar
to disk NoSF, with
$j_z$ decreasing over time, moving from an
almost purely positive valued distribution to a broader profile where
a substantial fraction of the clouds have a negative $j_z$. These
clouds rotate in the opposite sense to the galaxy as will be discussed
in section~\ref{sec:simtime_angmom}. This broadening is due to
interactions between clouds affecting their rotation. 

The diffuse heating once again moderates the evolution of the
cloud distributions, causing $j_z$ in disk SF+PEheat to show a smaller change over
time. While cloud interactions are still common (as we saw in
Figure~\ref{fig:mergers}), their effect on the
cloud's rotation is moderated by the denser, filamentary structure of
the surrounding ISM which maintains a low velocity dispersion as was seen in
Figure\ref{fig:diskevol}. This allows late forming clouds to form in
a more structured, less turbulent environment than without diffuse
heating, which plays a significant role in determining the rotation
of the clouds. 

The vertical distribution of the clouds is shown in the final panel,
Figure~\ref{fig:simtime_clouds_sf5}\,(f) and
Figure~\ref{fig:simtime_clouds_peheat}\,(f). The cloud scale height
grows over the course of the simulation in both disks, due to the
frequency of cloud collisions scattering clouds out of the galactic
mid-plane. With the exception of our last time analysis which has a
high degree of scatter, the scale height of the clouds is comparable
to that of the GMCs in the Milky Way at $\lesssim 35$\,pc
\citep{Stark2005}. Diffuse heating makes little difference to this
distribution, although we saw in Figure~\ref{fig:diskpdf}, that the
scale-height of the warm ISM is initially larger in disk SF+PEheat due to the added
pressure from the heating.

\subsubsection{Simulation Time: Velocity Dispersion vs Size Relation}
\label{sec:larson}

\begin{figure*}
\begin{center} 
\includegraphics[width=8cm]{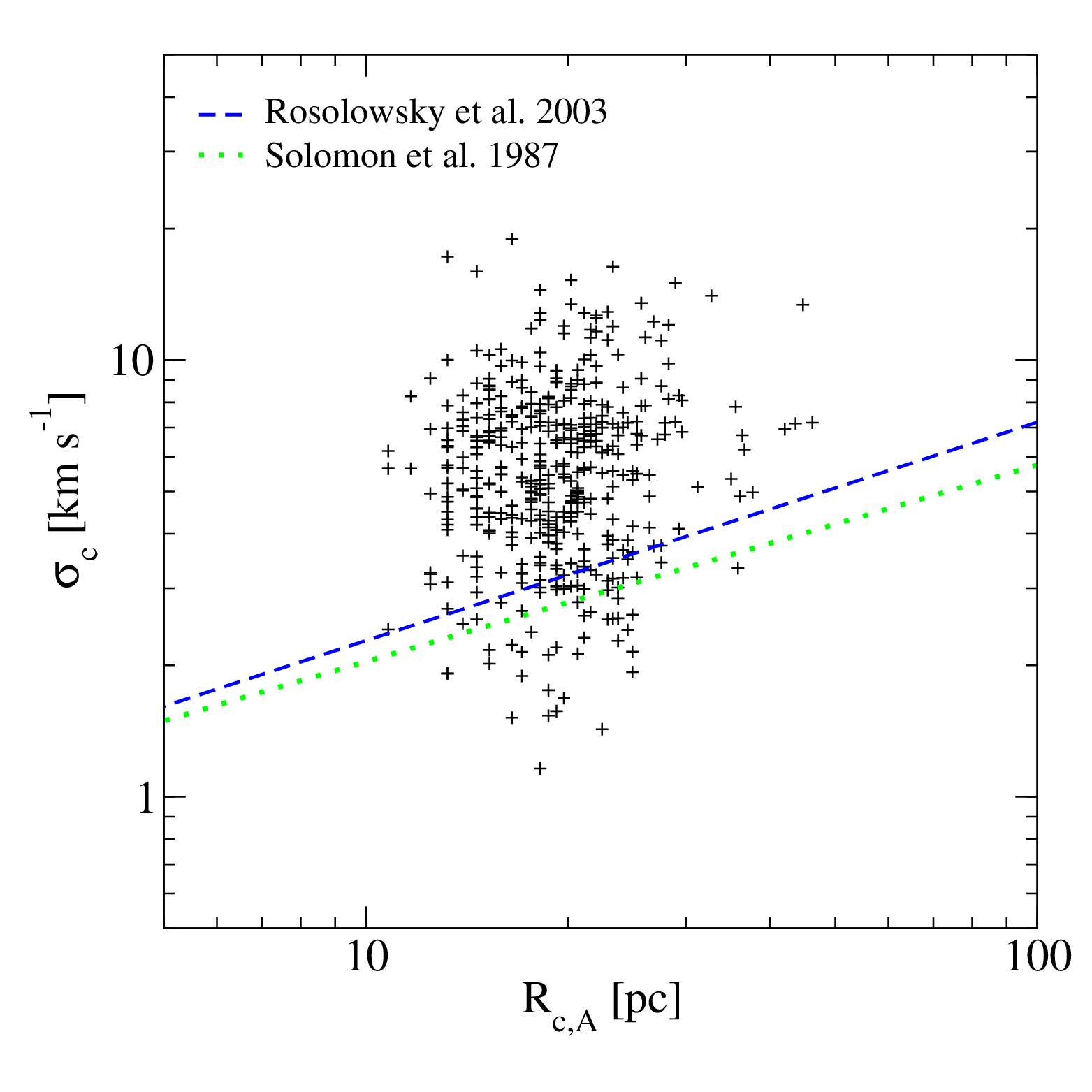}
\includegraphics[width=8cm]{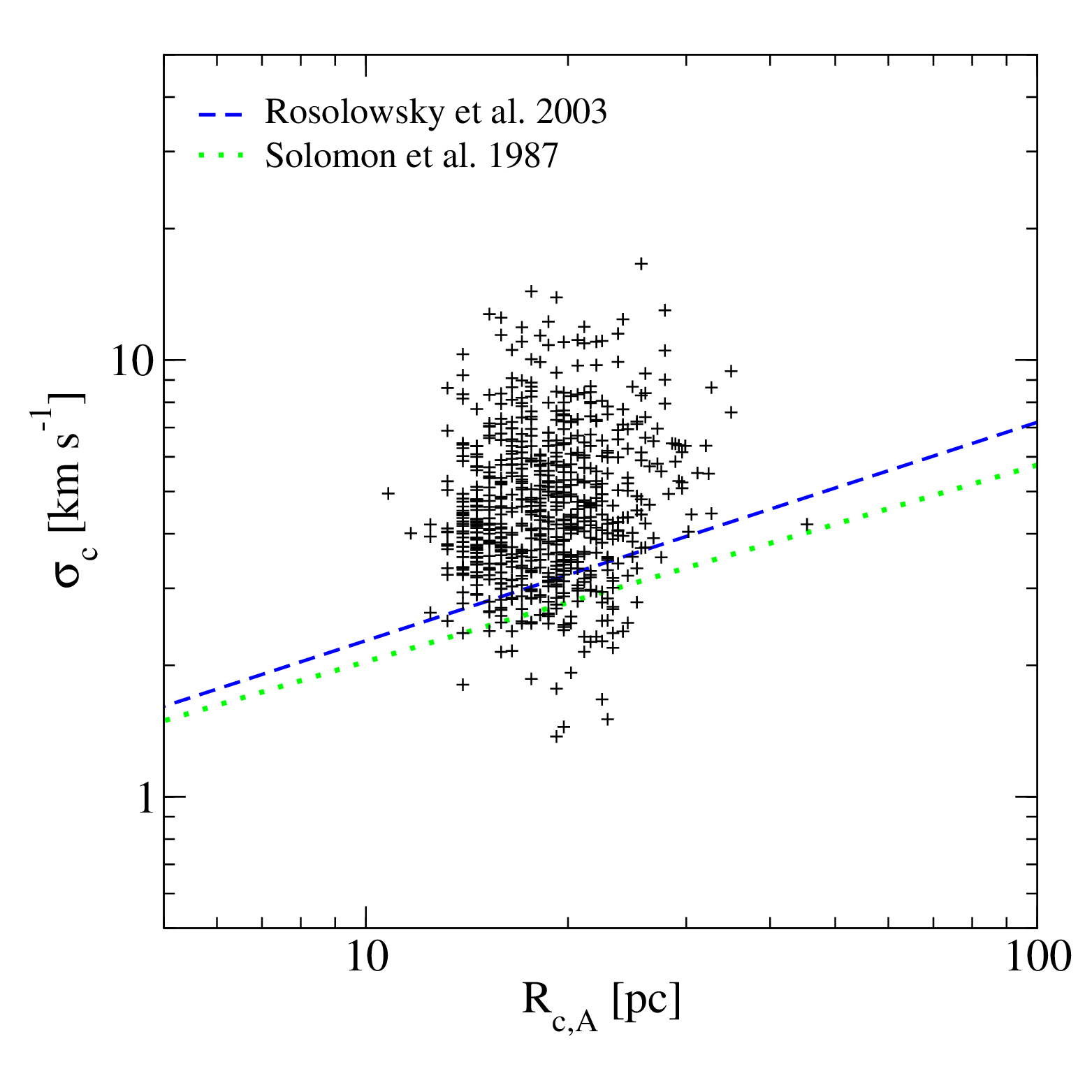}
\caption{Cloud velocity dispersion vs size relation for clouds
  with mass $M > 10^{5.5}$\,M$_\odot$ at t = 200\,Myr. Clouds in disk
  SFOnly are plotted in the left-hand pane and clouds
  in disk SF+PEheat are plotted on the
  right. $\sigma_c$ is the line of sight velocity of the clouds in the galaxy viewed at
  52$^\circ$ inclination angle, similar to our view of M33. The dashed
  blue line shows the result (slope and normalization) of the
  \citet{Rosolowsky2003} study of massive ($10^5$\,M$_\odot \lesssim M_c \lesssim
  10^6$\,M$_\odot$) GMCs in M33, with an exponent of $0.45 \pm
  0.02$. The green dotted line shows the result of the
  \citet{Solomon1987} study of Galactic GMCs with an exponent of $0.5
  \pm 0.05$. No obvious trend is seen in our results due to the limited
  range of radii of the clouds.
\label{fig:veldisp}}
\end{center} 
\end{figure*}

The internal cloud velocity dispersion plotted against cloud size is
shown for clouds present at t=200\,Myr in Figure~\ref{fig:veldisp} for
both disks. Only clouds with mass greater than $M >
10^{5.5}$\,M$_\odot$, whose internal structure has a reasonable chance
of being resolved, are plotted. (This cut-off is slightly lower ($M > 10^{5.5}$\,M$_\odot$) than
in TT09 ($M > 10^{6}$\,M$_\odot$) due to clouds being on
average more massive). $\sigma_c$ is the mass-averaged 1D internal velocity
dispersion of the clouds viewed at a $52^{\circ}$ inclination angle,
similar to our view of M33. 

Observed clouds in M33 \citep{Rosolowsky2003} and the Milky Way
\citep{Larson1981, Solomon1987} show a velocity dispersion (line
width) that increases as a power of their radii. The fit for these two
observed populations are also plotted in
Figure~\ref{fig:veldisp}. Unlike in disk NoSF, where we reproduced the
observed relation over an order of magnitude, neither of our disks
have a cloud population that shows a strong correlation between these
two quantities. This is likely due to the restricted range of radii
that the clouds now obtain, covering only a factor of four in range,
compared to in disk NoSF where they extended over a factor of 10 without the
inclusion of star formation to limit growth. 

In part, this uniformity of our current cloud population radii is due to our
initial conditions. We intentionally model clouds in an environment
typical to the Milky Way at the solar radius. The range in disk surface
density is only a factor of 10 as can be seen in
Figure~\ref{fig:diskevol} and decreases over time. This means that any
variation in cloud properties in our populations must come from their
interactions and internal physics, not from variations in the galactic
environment. The inclusion of energetic localized feedback might be
expected to ease this issue and create a wider range of cloud
radii. 

The effect of diffuse heating in disk SF+PEheat slightly reduces the spread in
$\sigma_c$, but otherwise has no impact. This is in difference to the
effect on the warm ISM, where the presence of diffuse heating
significantly reduced the velocity dispersion, as seen in
Figure~\ref{fig:diskevol}. The denser, self-gravitating cloud gas,
however, is less altered by this term.

\subsubsection{Simulation Time: Distribution of the Angular Momentum Vector}
\label{sec:simtime_angmom}

\begin{figure*}
\begin{center} 
\includegraphics[angle=270,width=\textwidth]{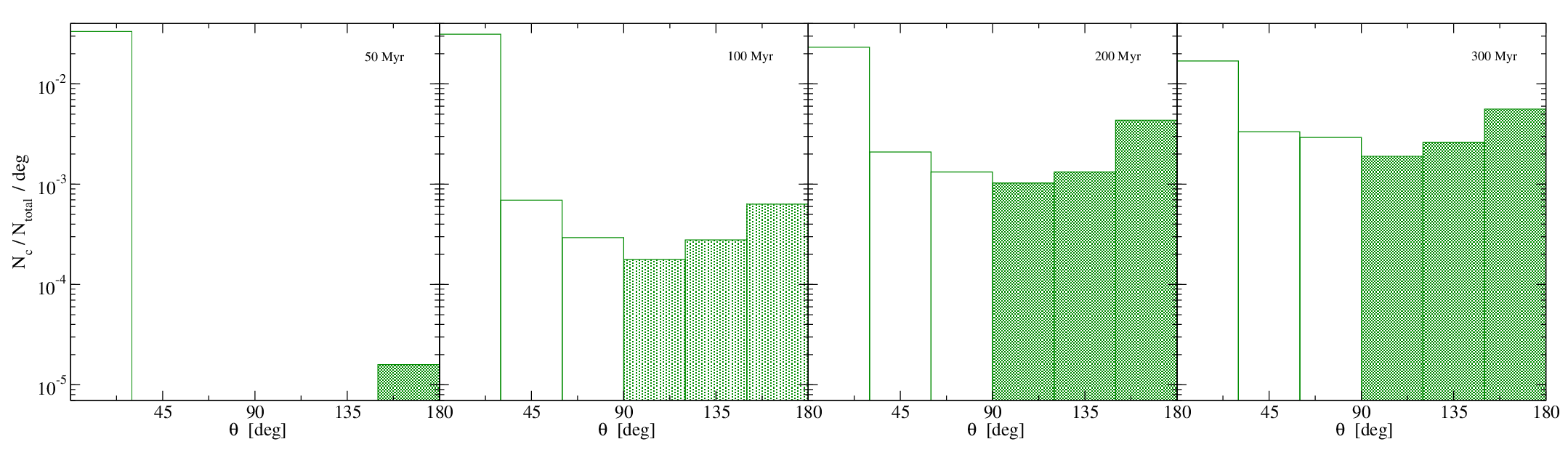}
\includegraphics[angle=270, width=\textwidth]{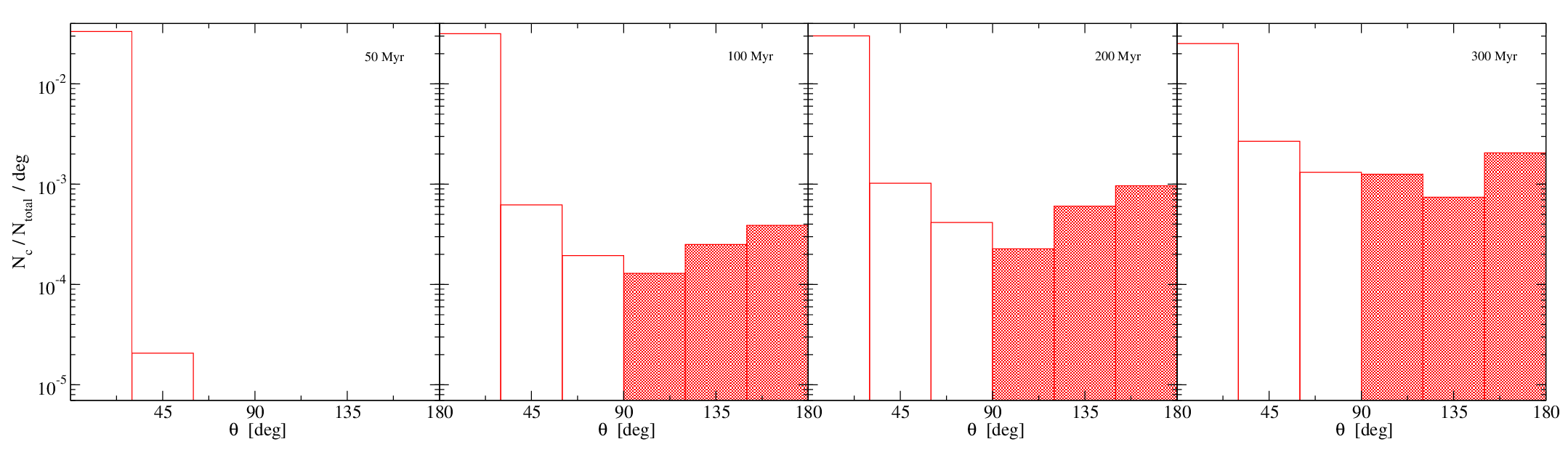}
\caption{Distribution of the angle between the cloud angular momentum
  vector and the galactic rotation axis at different times over the
  course of the simulation. The distribution for clouds in disk SFOnly is shown
  in the top panel and in disk PEheat in the bottom
  panel. The shaded bars indicate retrograde
  motion. The presence of diffuse heating significantly decreases the
  number of retrograde clouds at late times. By 300\,Myr, the number of retrograde
  clouds has reached 30\% in the disk without diffuse heating and 12\% in the disk
  with heating.}
\label{fig:angmom}
\end{center} 
\end{figure*}

The distribution of angles, $\theta$, between the cloud's own angular
momentum vector and the galactic rotation axis is shown in
Figure~\ref{fig:angmom} for both disks at four simulation times; 50,
100, 200 and 300\,Myr. $\theta$ for clouds in disk SFOnly are shown on
the top panel while $\theta$ for the clouds in Disk SF+PEheat
 are in the lower panel. Clouds with $0 < \theta < 90$
rotate about their center of mass in the same sense as the galaxy's
angular momentum vector and are considered to be {\it prograde}
rotators. Clouds with $90 < \theta < 180$ rotate in the opposite
sense to the galaxy and are {\it retrograde} rotators. As the first
clouds begin to form in the disk, we see an almost purely prograde
population for both simulations. These clouds predominantly feel only
the disk sheer as cloud-cloud interactions are still low in the partially
fragmented disk. As the simulation continues, a retrograde population
of clouds develops and by 300\,Myr, 30\% of the clouds in the disk
without diffuse heating rotate in the opposite sense to the
galaxy. Clouds present at these later times feel not only the sheer of
the disk, but also the gravitational pull from neighboring clouds,
including the effect of mergers. While forming in this more complex
environment, clouds can develop a retrograde motion, or be later scattered by
neighboring clouds to switch their sense of rotation. 

When diffuse heating is included in disk SF+PEheat, the fraction of clouds that develop
a retrograde motion is significantly reduced, with only $12$\%
rotating in a counter sense to the galaxy at 300\,Myr. That this was
true was indicated in Figure~\ref{fig:simtime_clouds_peheat}\,(e),
where the specific angular momentum profile contained only a small
component with $j_z < 0.0$. As mentioned in
section~\ref{sec:simtime_distributions}, this is the result of the
denser warm ISM, whose filamentary structure acts to encourage the
clouds to remain prograde.

The effect of the ISM environment on cloud properties was also noted
by \citet{Dobbs2008} in their single fluid model of the ISM. In their disk
simulation using an isothermal warm gas, no clouds arose with
retrograde rotation, a fact they attribute to the absence of a clumpy
medium required to cause collisional interactions between clouds. In our model,
the ISM is fully multiphase, but the diffuse heating has increased the
density and structure of the warm gas. We do not find a decrease in
the number of cloud mergers, but similarly to \citet{Dobbs2008}, we
find the warm ISM is playing a significant role in determining the
cloud rotation. 

The result that the disk environment can impact the cloud properties
is particular interesting, since that was not a previous prediction
from analytical models or observations. In disk NoSF, the
percentage of clouds that develop a retrograde
motion was equal to disk SFOnly, supporting the claim that it is the difference in the warm
ISM that controls the development of retrograde motion. 

Observations by \citet{Rosolowsky2003} find a 40\% split
between prograde and retrograde clouds in the sample of GMCs in
M33. Both our disks support the hypothesis that interactions between
self-gravitating clouds are responsible for the distribution in angular rotation.

\subsection{GMC Properties with Cloud Age}
\label{sec:cloudtime}

\begin{figure} 
\begin{center} 
\includegraphics[width=7cm]{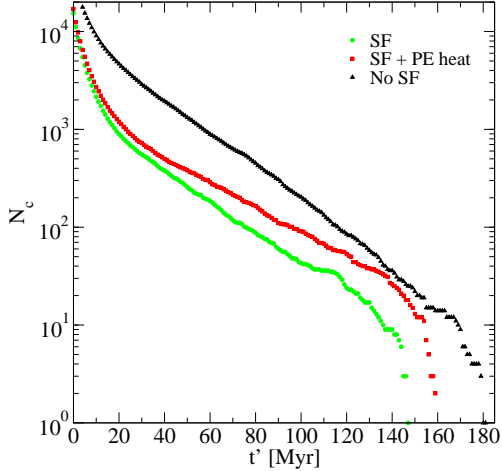}
\caption{Distribution of cloud lifetimes in the simulations. Green
  circles show the clouds in disk SFOnly. Red squares depict clouds in
  disk SF+PEheat and black triangles show the results from disk
  NoSF. The majority of clouds live
  only a few Myr with only of order 0.5\% reaching 100\,Myr. This high
  end tail is probably due to the lack of energetic feedback.
\label{fig:cloudlifetimes}}
\end{center} 
\end{figure}

By tracking the clouds through the simulation over time, as described
in section~\ref{sec:ic}, we are able to compare properties of clouds
that have the same age, $t'$, but exist at different simulation
times. To avoid our results being affected by the initial
fragmentation of the disk (during which cloud interactions are
low) we only consider clouds born after 140\,Myr in our analysis.

The spread of ages in the clouds of our simulations are shown in
Figure~\ref{fig:cloudlifetimes}. In the two simulations presented in
this paper, the vast majority of clouds only live
between $t'_{\rm age} = 0 - 20$\, Myr. This number
drops by half after 3\,Myr and by a factor of ten by 20\,Myr. Of
the 7\% that live longer than this, only 0.5\% survive to
100\,Myr. 

Without star formation, the decrease in cloud number is steady with
age, as cloud destruction occurs purely as a function of the merger
rate which is approximately constant. The addition of star formation
increases the cloud mortality rate by a factor of 3-4 by a cloud age
of 10\,Myr, having the greatest impact on the younger, less massive
clouds (see below for details of the evolution in the cloud mass). It
is likely these very young objects are forming close to other clouds
who they almost immediately merge with, or that their smaller size
causes them to be disrupted quickly by star formation. 

Including the diffuse heating in disk SF+PEheat decreases the star formation rate and allows
clouds to live longer. It has the most effect on older clouds, whose
higher mass makes them more resilient to mergers and hence their death
is controlled primarily by their star formation. 

The lifetimes of GMCs is still a matter of intense debate. A current
prevailing point of view, however, is that GMCs live 1-2 dynamical
times, putting their age in the range of 5-20\,Myr \citep{McKee2007,
  Murray2010}. This agrees well with our clouds, a surprising
result since we include no form of energetic feedback which has been
considered to be one of the main factors in cloud destruction. 

\subsubsection{Cloud Time: Cloud Property Distributions}
\label{sec:cloudtime_distributions}

\begin{figure*} 
\begin{center} 
\includegraphics[width=15cm]{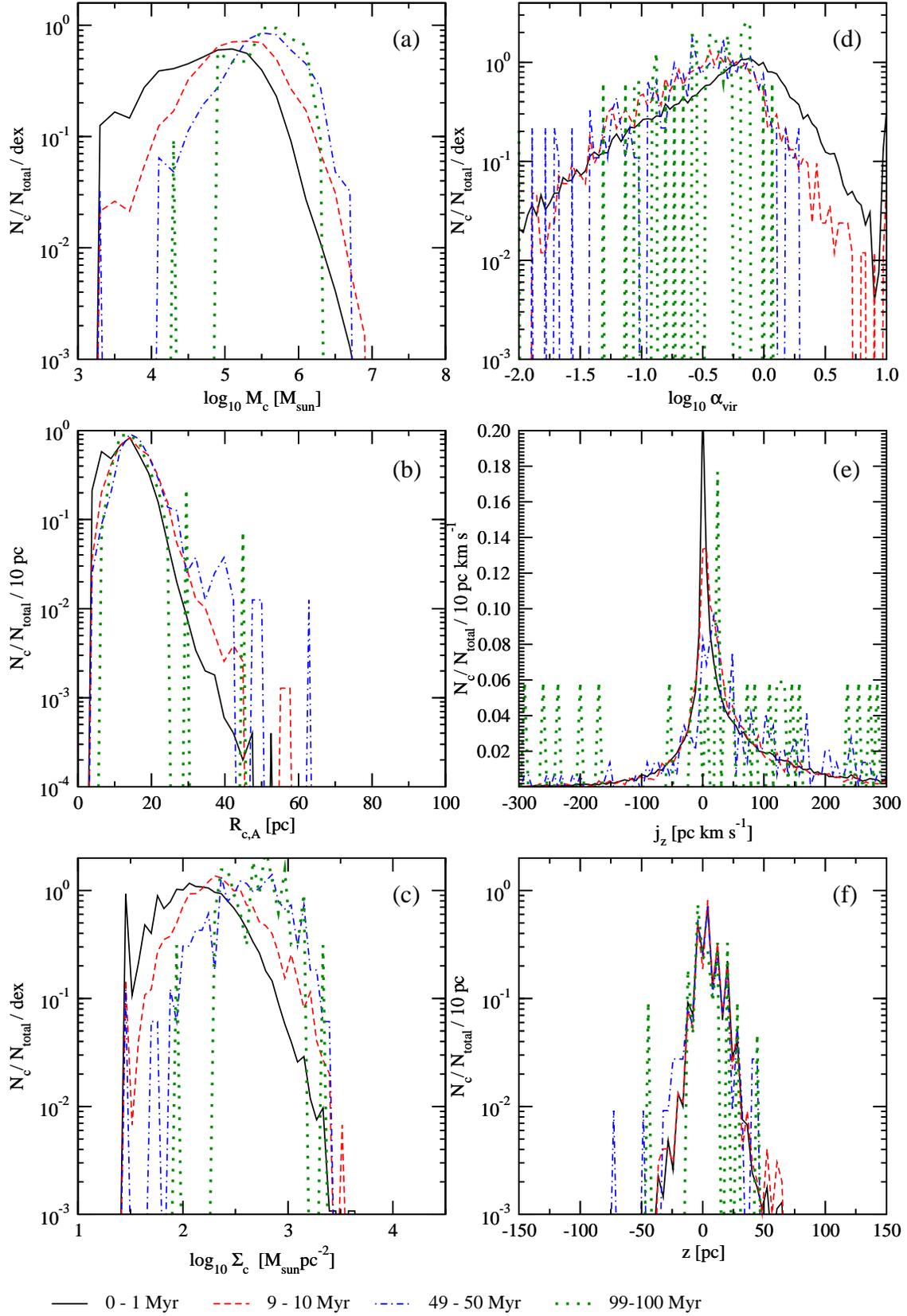}
\caption{Normalized distributions of the GMC properties described
  in Figure~\ref{fig:simtime_clouds_sf5} for disk SFOnly, but now showing results for
  different cloud ages: 0-1\,Myr (solid lines), 9-10\,Myr (dashed
  lines), 49-50\,Myr (dot-dashed lines) and 99-100\,Myr (dotted
  lines). All clouds in this analysis were born after 140\,Myr of disk
  evolution, i.e. in the fully fragmented phase. Top left, (a): cloud
  mass, $M_c$. Middle left, (b): cloud radius, $R_{C,A}$. Bottom left,
  (c): mass surface density, $\Sigma_c$. Top right, (d): virial
  parameter, $\alpha_{\rm vir}$. Middle right, (e): vertical component
  of the specific angular momentum, $j_z$. Bottom right, (f): cloud
  center-of-mass vertical positions, $z$.}
\label{fig:cloudtime_sf5}
\end{center} 
\end{figure*}

\begin{figure*} 
\begin{center} 
\includegraphics[width=15cm]{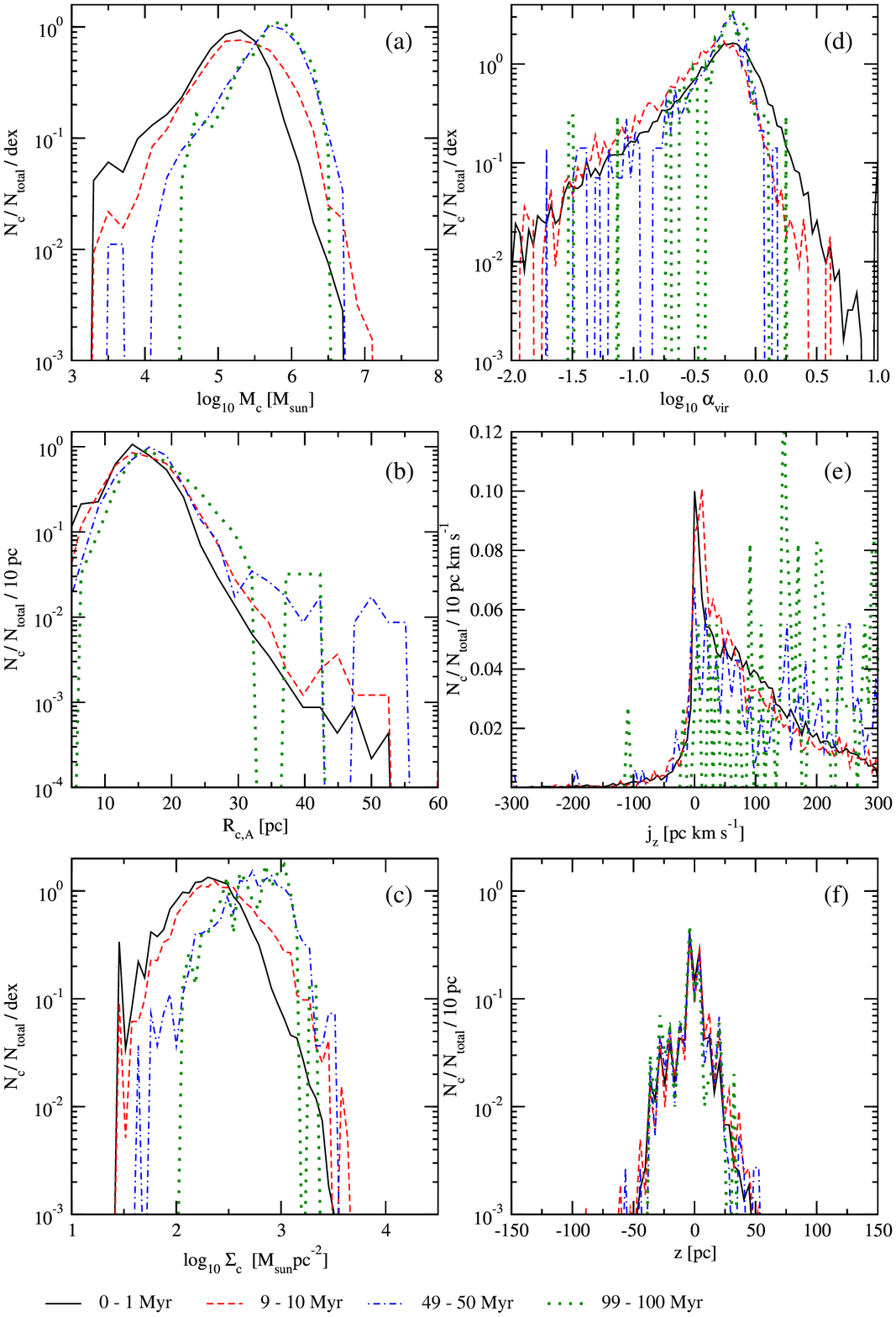}
\caption{Normalized distributions of the GMC properties, as described
  in Figure~\ref{fig:simtime_clouds_peheat} for Disk SF+PEheat, for
  cloud ages: 0-1\,Myr (solid lines), 9-10\,Myr (dashed
  lines), 49-50\,Myr (dot-dashed lines) and 99-100\,Myr (dotted lines)
for clouds born after 140\,Myr of disk evolution, i.e. in the fully
fragmented phase. Top left, (a): cloud mass, $M_c$. Middle left, (b): cloud
radius, $R_{C,A} \equiv (A_c/\pi)^{1/2}$. Bottom left, (c): mass
surface density, $\Sigma_c$. Top right, (d): virial parameter,
$\alpha_{\rm vir}$. Middle right, (e):
vertical component of the specific angular momentum, $j_z$. Bottom
right, (f): cloud center-of-mass vertical positions, $z$.}
\label{fig:cloudtime_peheat}
\end{center} 
\end{figure*}

The properties of the clouds as a function of their age are shown in
Figures~\ref{fig:cloudtime_sf5} (disk SFOnly) and
\ref{fig:cloudtime_peheat} (disk SF+PEheat) for
the same distributions plotted in Figure~\ref{fig:simtime_clouds_sf5}
and \ref{fig:simtime_clouds_peheat}; cloud mass, radii,
surface density, virial parameter, vertical component of the angular
momentum and their vertical distribution. The solid black line shows
recently formed clouds that are less than 1\,Myr old. The red dashed
line shows the distributions for clouds between 9-10\,Myr
old, the estimated observed age for a GMC. The blue dot-dashed line is
for cloud ages 49-50\,Myr and the green dotted line for clouds
99-100\,Myr old. These two last lines show the properties of the older
clouds in the simulation although, as shown in Figure~\ref{fig:cloudlifetimes},
significantly fewer clouds live to these ages, with only 44 clouds
forming the distribution of the 99-100\,Myr profiles in disk SFOnly
and 92 clouds in disk SF+PEheat. 

In both the disks, we see that older clouds are on average more
massive, have larger radii, higher surface density and are slightly
more gravitationally bound. These trends are more pronounced than in
disk NoSF since the gas depletion from star formation destroys a large
number of younger clouds, as seen in
Figure~\ref{fig:cloudlifetimes}. Clouds with masses
$M \lesssim 10^5$\,$_\odot$ are particularly prone to early mortality
through star formation and mergers. The clouds that survive their early
years are more massive, accumulating gas through accretion and mergers
at a faster rate than their star formation can deplete it. This effect
is most marked in disk SFOnly. In disk SF+PEheat when diffuse heating is
present, less clouds with masses $\lesssim 10^5$\,M$_\odot$ are born
due to the additional heating increasing the Jeans length, raising
the size of an object formed through gravitational collapse. This
reduces the evolution of the low mass distribution at in the first
10\,Myr of the cloud's life. 

In contrast to this, the distributions of vertical angular momentum,
Figure~\ref{fig:cloudtime_sf5}\,(e) and Figure~\ref{fig:cloudtime_peheat}\,(e), and vertical position above the
disk, Figure~\ref{fig:cloudtime_sf5}\,(f) and
Figure~\ref{fig:cloudtime_peheat}\,(f), are largely independent of
cloud age. Without diffuse heating, newly formed clouds in disk SFOnly have a very
low vertical angular momentum, causing the distribution to be sharply
peaked. Cloud interactions later broaden this profile slightly and
reduce the peak, but the form remains largely unchanged. When diffuse
heating is present in disk SF+PEheat, the clouds are born with a wider range of positive
(prograde) angular momentum values that also show little change
over time. The filamentary structure of the ISM has produced a more
dominantly prograde cloud population as seen in
Figure~\ref{fig:angmom}, and the higher values of $j_z$ means the
probability of later producing a retrograde cloud in a cloud collision is low.\\*[0.3cm]

\subsubsection{Cloud Time: Distribution of the Angular Momentum Vector}

\begin{figure*} 
\begin{center} 
\includegraphics[angle=270,width=\textwidth]{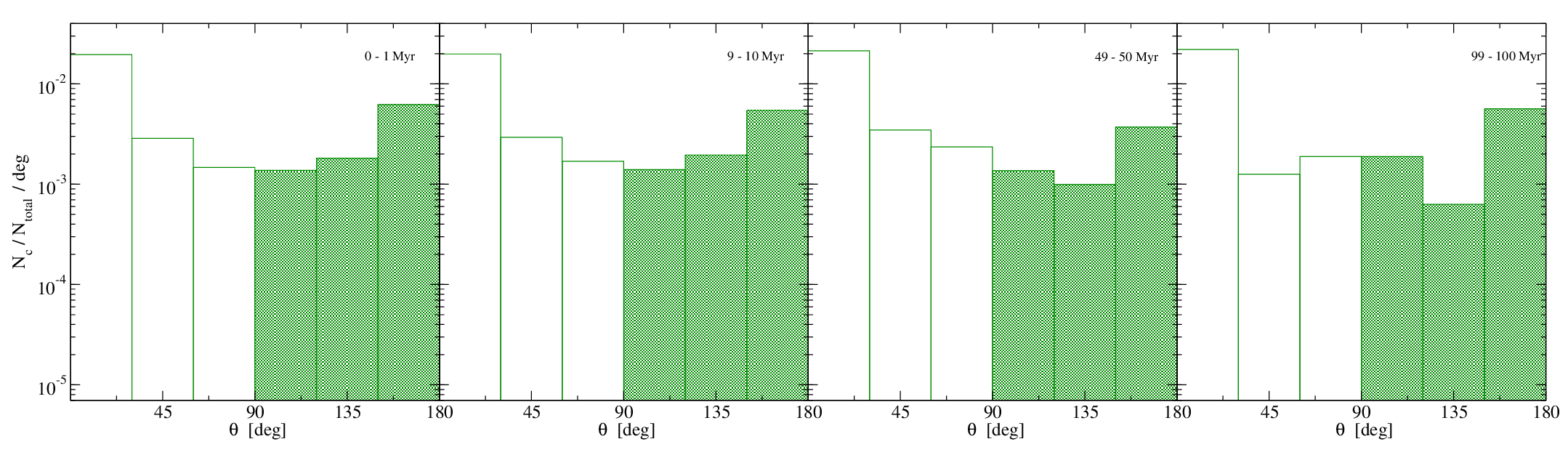}
\includegraphics[angle=270, width=\textwidth]{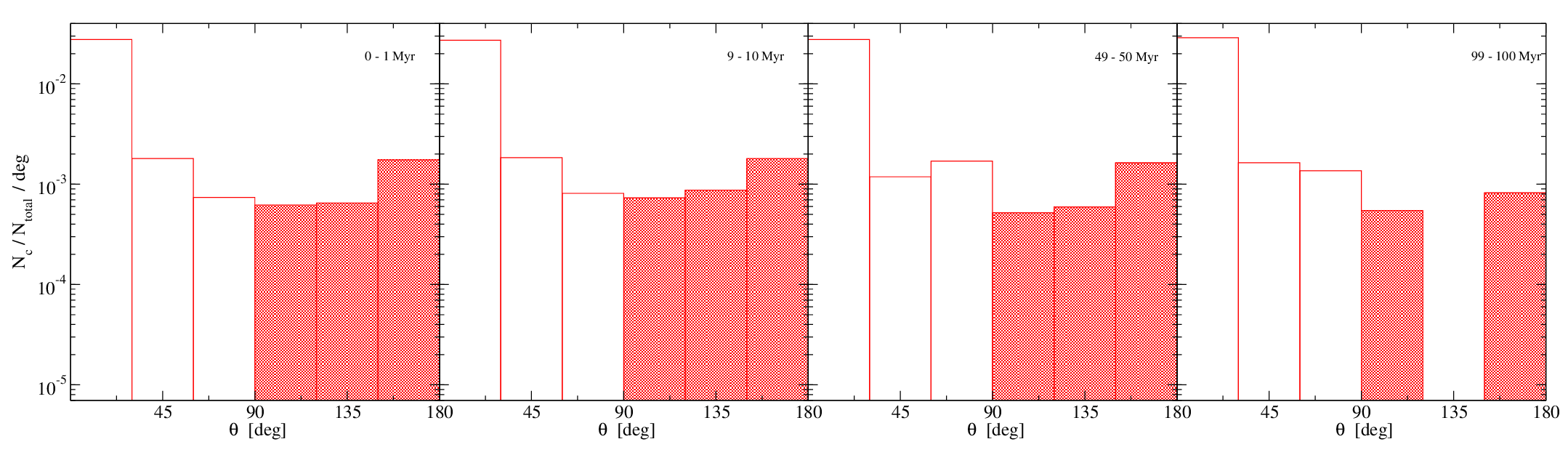}
\caption{Distribution of the angle between the cloud angular momentum
  vector and the galactic rotation axis for clouds of different ages,
  born after 140\,Myr, 
  in disk SFOnly (top) and disk SF+PEheat (bottom). The shaded bars indicate retrograde
  motion. The percentage of clouds rotating retrograde in the top
  row is 28\%, 26\%, 18\%, 25\% respectively for each panel and 9\%, 10\%,
  8\%, 4\% for the bottom row.
\label{fig:angmom_cloudt}}
\end{center} 
\end{figure*}

The robustness of the angular momentum with respect to cloud age is
shown in the distribution of angles, $\theta$, between the cloud
angular momentum vector and the galactic rotation axis plotted in
Figure~\ref{fig:angmom_cloudt}. Clouds at four different ages are
shown; newly born clouds with $t' < 1$\,Myr, then clouds with ages
$t' = 9-10$\,Myr and $49-50$\,Myr and finally our oldest clouds with
$t' = 99-100$\,Myr. The top panel shows the distribution of $\theta$
for disk SFOnly while the lower panel shows the
distribution for clouds in disk SF+PEheat. As with
Figure~\ref{fig:angmom}, the shaded bars show angles that equate to a
retrograde rotation.

The percentage of retrograde rotating clouds remains about 25\% for
all age ranges in disk SFOnly and about 10\% in disk SF+PEheat. Since
all these clouds are born after
140\,Myr in the simulation, they feel the forces from the fully
fragmented disk, in addition to the sheer. The gravitational
interactions from neighboring clouds on the newly forming bodies cause
the retrograde population to form. As we can see from the first panel
of Figure~\ref{fig:angmom}, clouds forming in pristine gas are always prograde. 

As we have seen previously, the effect of diffuse heating is to reduce
the fraction of retrograde clouds, due to the filamentary structure of
the warm ISM having a strong impact on the newly forming clouds. 

\section{Star Formation}
\label{sec:stars}

\begin{figure} 
\begin{center} 
\includegraphics[width=8cm]{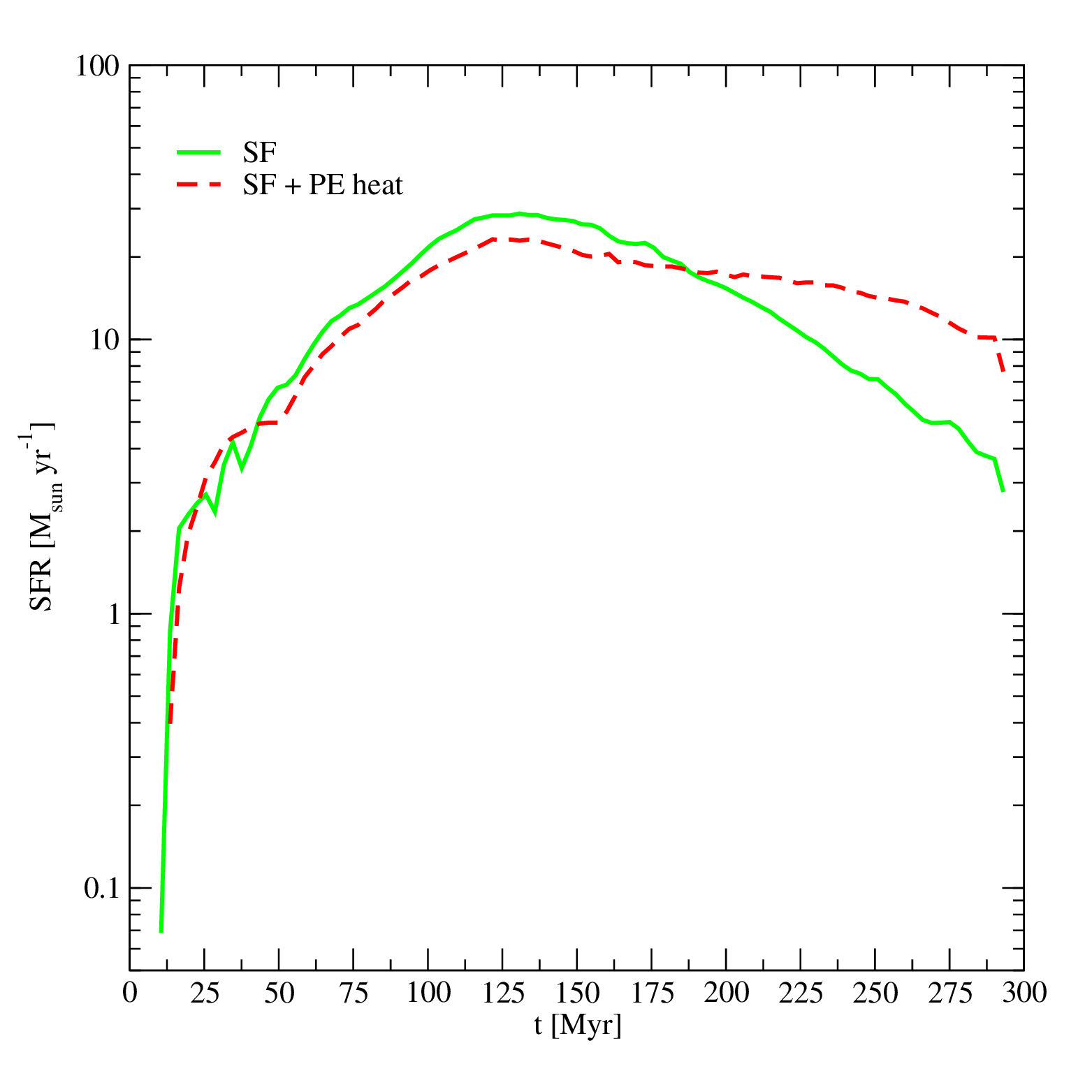}
\caption{The star formation rate over the course of the simulation for
  disk SFOnly (green solid line) and SF+PEheat (red
  dashed line). Like the cloud population, the SFR peaks
  shortly after the disk has become fully fragmented. As gas is
  converted into stars and depleted from the clouds, it begins to
  drop. This decrease in SFR is significantly slower when
  diffuse heating is included, dropping to a value of
  10\,M$_{\odot}$\,yr$^{-1}$ rather than  3\,M$_{\odot}$\,yr$^{-1}$
  when diffuse heating is not included. Around 200\,Myr, before gas
  depletion has removed most of the gas in the clouds with no diffuse
  heating, the SFR is greater than the estimated Milky Way value by a
  factor of 10. This is probably due to the lack of localized feedback.
\label{fig:sfr_history}}
\end{center} 
\end{figure}

\begin{figure*} 
\begin{center} 
\includegraphics[width=8cm]{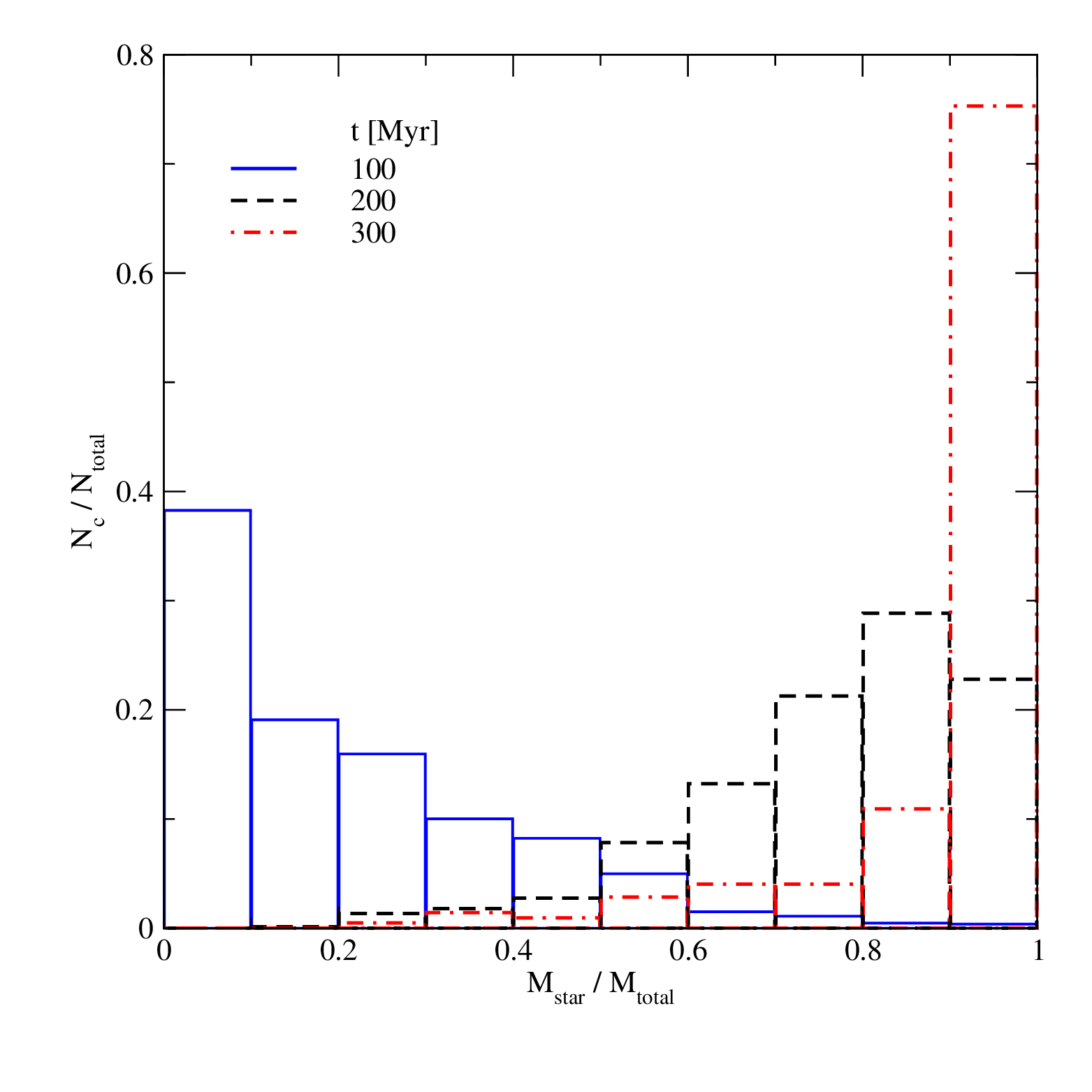}
\includegraphics[width=8cm]{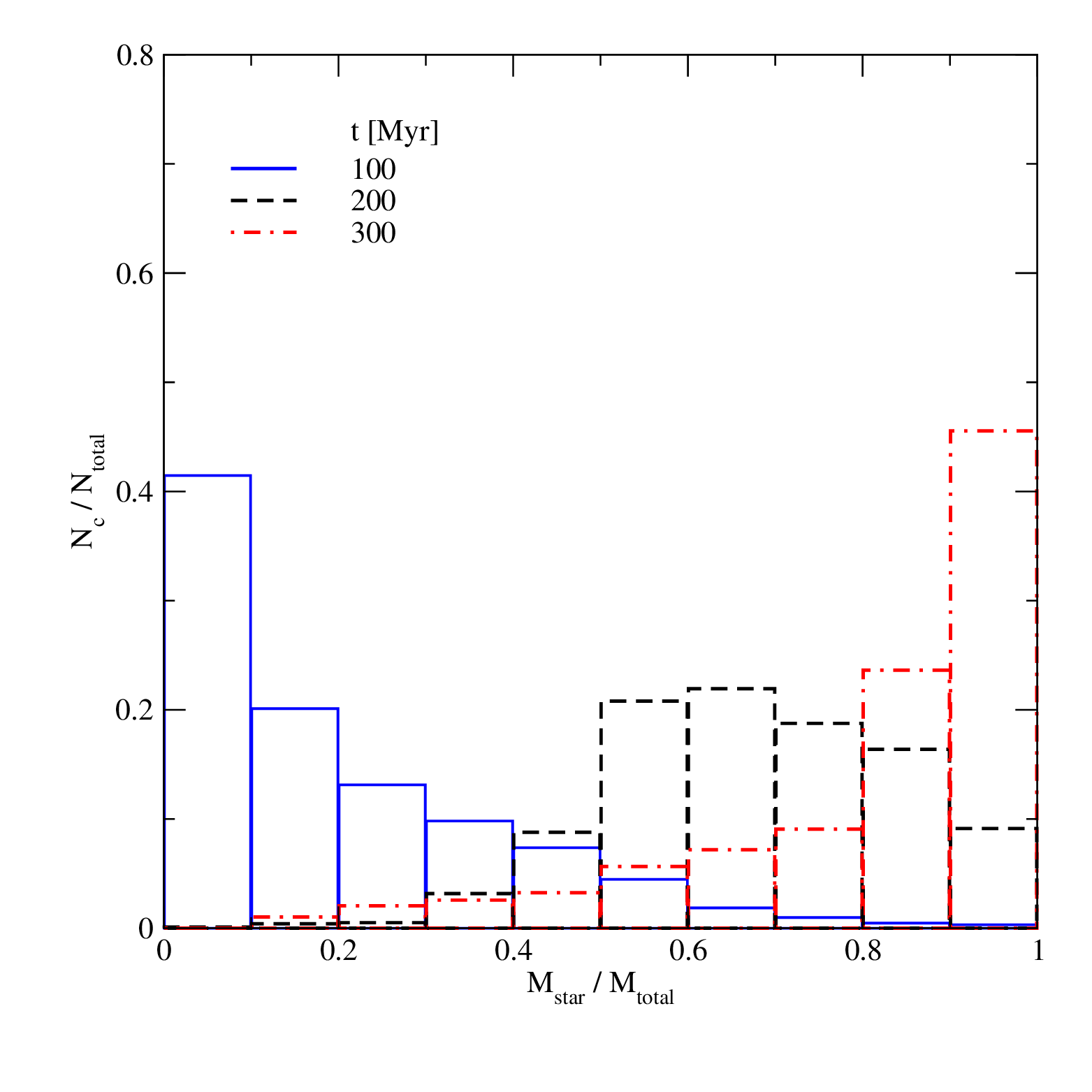}
\caption{Fraction of mass in stars for clouds at simulation times
  100\,Myr (solid blue line), 200\,Myr (dashed black line) and
  300\,Myr (red dot-dashed line). Left plot shows results for
  disk SFOnly while the right-hand plot shows
  the clouds in disk SF+PEheat.  
\label{fig:star_frac}}
\end{center} 
\end{figure*}

\begin{figure} 
\begin{center} 
\includegraphics[width=8cm]{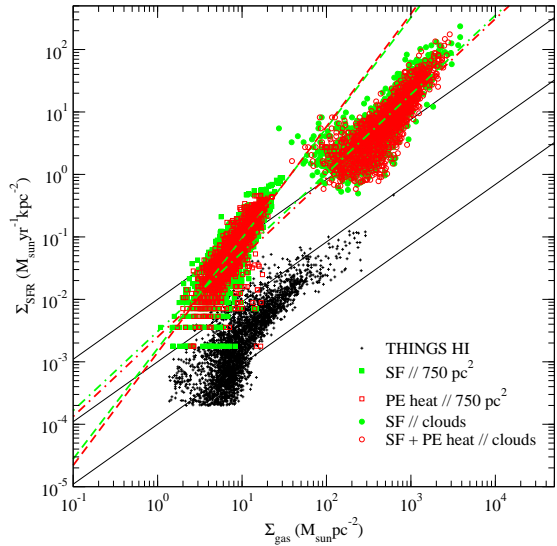}
\caption{The surface SFR, $\Sigma_{\rm sfr}$, vs. surface gas density,
  $\Sigma_{\rm gas}$ for the cloud populations in our two
  disks at t = 200\,Myr. The surface area is taken to be in the
  y-z plane, i.e. as if the clouds were viewed from inside the
  disk. Filled green squares are for an averaged area of 750\,pc across
  around each cloud (equivalent spatial resolution to the
  observations) in disk SFOnly with a
  best fit gradient of $\alpha_{\rm sfr} = 1.77$. Open red squares
  show the same averaged area around clouds in disk SF+PEheat with $\alpha_{\rm sfr} = 1.81$. Filled green
  circles show the relation averaged over
  just the cloud's surface area in disk SFOnly, with an $\alpha_{\rm sfr} =
  1.27$. Open red circles show the same in disk SF+PEheat and with
  an $\alpha_{\rm sfr} = 1.27$. Black crosses display the 
  observational results from the THINGS survey
  \citep{Bigiel2008}. The diagonal solid lines mark constant star
  formation efficiency, indicating the level of $\Sigma_{\rm SFR}$
  required to consume 1\%, 10\% and 100\% of the gas in $10^8$\,Myr
  (as shown in \citet{Bigiel2008}). When averaged over the same area
  as the observations, our SFR is a factor of $\sim 10$ too high,
  probably due to the lack of localized feedback. At this resolution,
  we would expect our gas to be largely atomic, with a steeper
  gradient similar to that of the observations below $\Sigma_{\rm gas} \lesssim
  10$\,M$_\odot$\,pc$^{-2}$. At the cloud
  resolution level, the gas should be at least 50\% molecular
  (depending on the size of the atomic envelope) and has a gradient
  approaching the observations for pure molecular gas,
  $\Sigma_{\rm gas} \gtrsim 10$\,M$_\odot$\,pc$^{-2}$.
\label{fig:schmidt}}
\end{center} 
\end{figure}

\begin{figure*} 
\begin{center} 
\includegraphics[width=8cm]{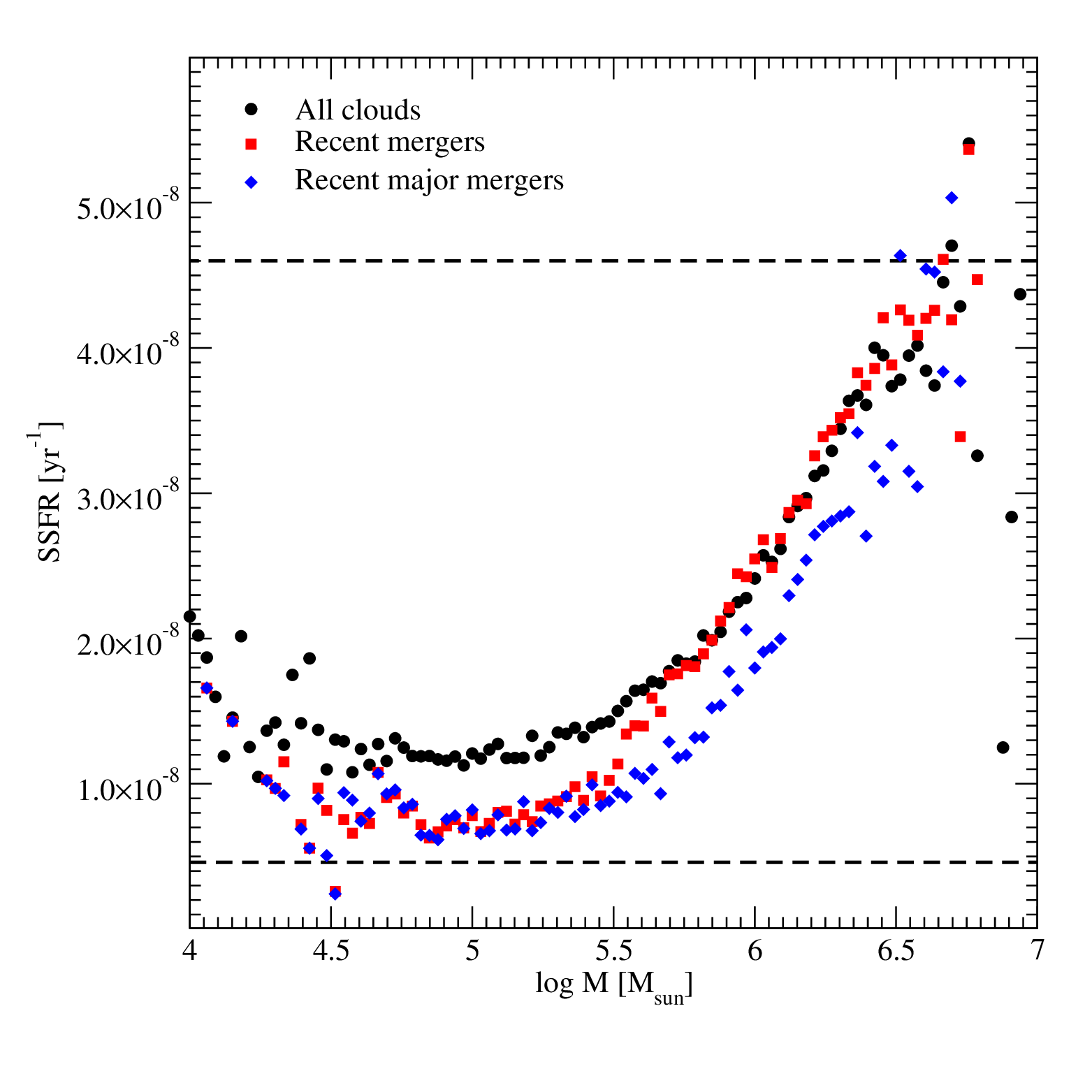}
\includegraphics[width=8cm]{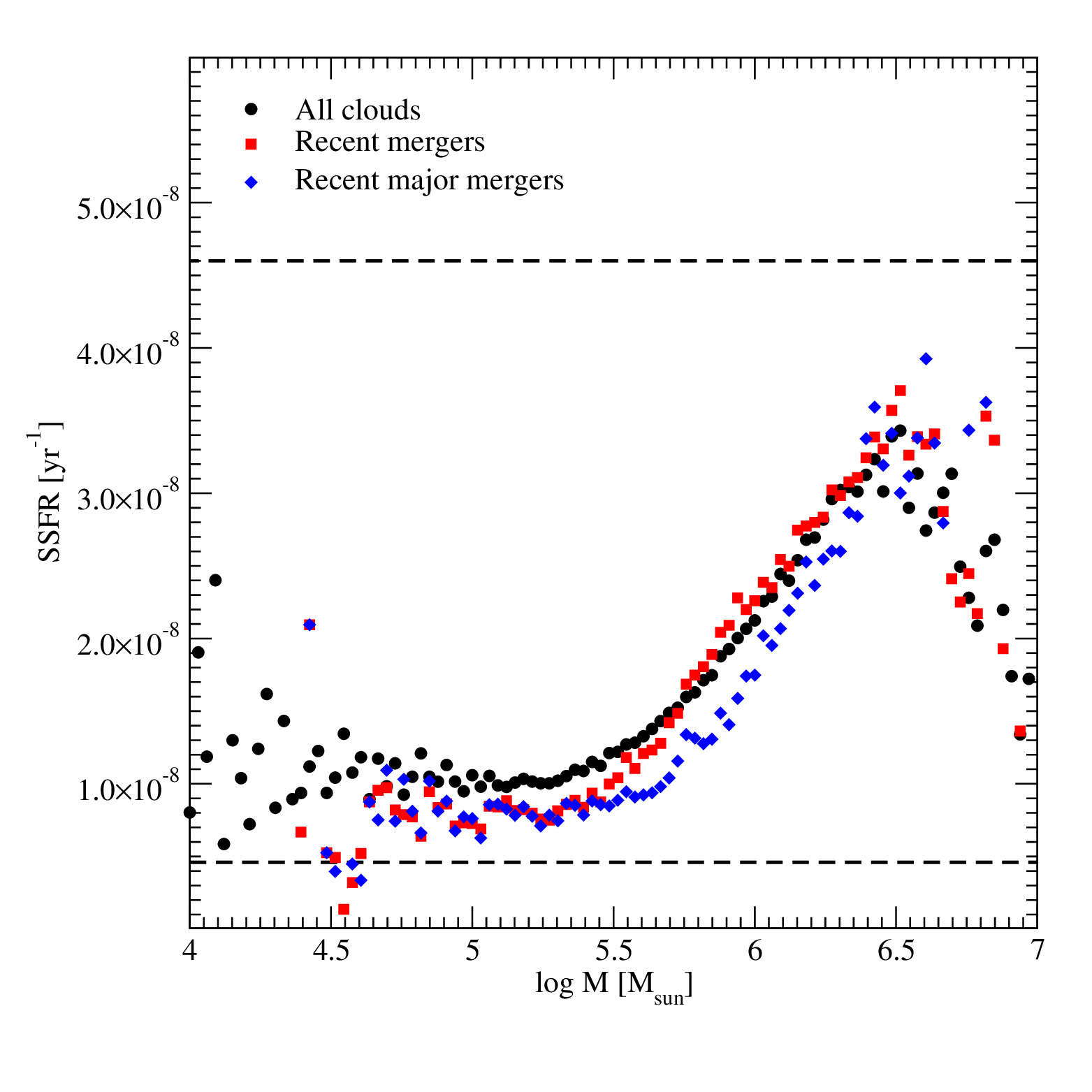}
\caption{Specific star formation rate averaged over one dynamical
  time (cloud free-fall time) vs. cloud gas mass for individual
  clouds in disk SFOnly (left plot) and disk SF+PEheat (right). Black
  circles show the distribution for all
  clouds, red squares include only clouds that have undergone a merger in the last dynamical time and
  blue diamonds include only the clouds for whom that merger was with a cloud
  greater than 50\% of its mass. 
\label{fig:ssfr}}
\end{center} 
\end{figure*}

The star formation history for the disk is plotted in
Figure~\ref{fig:sfr_history}. As with the cloud formation history in
Figure~\ref{fig:formation_history}, the star formation initially
increases steeply as the disk fragments. It reaches a peak value
around 125\,Myr and then declines steadily as gas is consumed in the
disk. 

At 200\,Myr, the star formation rate is 15\,M$_\odot$yr$^{-1}$ for
disk SFOnly and 17\,M$_\odot$yr$^{-1}$ for disk SF+PEheat. By the end
of the simulation, this has
dropped to 3\,M$_\odot$yr$^{-1}$ and 10\,M$_\odot$yr$^{-1}$,
respectively, although the lower numbers are purely a factor of the
gas depletion. The Milky Way is estimated to have a star formation
rate of order $1-3$\,M$_\odot$yr$^{-1}$ \citep{Murray2010a,
  Williams1997}. The fact we are higher than this though, is not
surprising due to our lack of localized feedback.

The addition of diffuse heating in disk SF+PEheat initially reduces the fraction of
dense cloud gas in the disk, as we saw in Section~\ref{sec:ism},
causing the SFR to be lower over the first $\sim 175$\,Myr than when
heating was not included. This causes the gas to be depleted in the
disk at a slower rate, resulting in more gas being available at later
times. This can be seen in Figure~\ref{fig:star_frac} which shows the
distribution of the fraction of mass in stars for the clouds
present at three different simulation times. Early in the simulation,
clouds are gas dominated with a low ($< 0.1$) stellar fraction. By
300\,Myr, a large fraction of the gas has converted into stars, making
the majority of the clouds strongly stellar dominated. This is
especially true for clouds in disk SFOnly, where
$3/4$ of the clouds have a stellar fraction $> 0.9$ by the end of the
simulation. This inevitably causes the star formation to decrease and
after 175\,Myr, the star formation rate in disk SFOnly has dropped
below the rate in disk SF+PEheat because the
gas abundance has become too low.  

The Kennicutt-Schmidt relation is an empirical measurement of the
relationship between the surface SFR, $\Sigma_{\rm sfr}$, and the
surface gas density, $\Sigma_{\rm gas}$, in a galaxy. It describes how
efficiently a galaxy is converting gas into stars and takes the form
shown in Equation~\ref{eq:ks}. The value of the exponent, $\alpha_{\rm
  sfr}$, has been measured by several groups
\citep[e.g.][]{Kennicutt1998, Wong2002, Bigiel2008} for both the Milky
Way and other galaxies and found it to vary between $\alpha_{\rm
  sfr} \approx 1-3$. Most recent work \citep{Bigiel2008, Wong2002}
suggests that the correlation is truly with molecular gas,
$\Sigma_{\rm H_2}$, rather than total gas, $\Sigma_{\rm gas} =
\Sigma_{\rm HI}+\Sigma_{\rm H_2}$, for which $\alpha_{\rm sfr, H_2} = 1.0\pm
0.2$. When averaged with the atomic component, the exponent
increases before the relation breaks down around $\Sigma_{\rm gas}
\approx 10$\,M$_\odot$pc$^{-2}$ in the outer regions of the disk where
the gas becomes saturated with HI. 

Figure~\ref{fig:schmidt} shows this relation plotted for our
disks at $t = 200$\,Myr. The diagonal lines mark constant star formation
efficiency and the black crosses show the observations from The
HI Nearby Galaxy Survey (THINGS) \citep{Bigiel2008}. For these
observations, the fall-off around $10$\,M$_\odot$pc$^{-2}$ is due to
the dominance of atomic gas, with the points at higher densities with
$\alpha_{\rm sfr} = 1.0 \pm 0.2$ being purely from molecular $H_2$. The results
for disk SFOnly are shown by green filled
symbols while disk SF+PEheat is marked in
red open symbols. Square symbols show the result of averaging over an
area of $750$\,pc across around each cloud, the equivalent spatial
resolution to the THINGS survey. The circles show the quantities
averaged over the individual clouds themselves, with area of order
15\,pc across. 

For equivalent spatial resolution (squares) we see that our star
formation rate is higher than the observations by a factor of 10 in
both our simulations. This agrees with what we saw in
Figure~\ref{fig:sfr_history} and is likely due to not having a source
of localized feedback. The gradient here is $\alpha_{\rm sfr} 1.77$ for
the disk without diffuse heating and $1.81$ for when heating is
included. This is steeper than the fit for the molecular gas found by
\citet{Bigiel2008}, but in closer agreement with the gradient of the
atomic gas. Given the size of the region we are averaging over,
compared to our cloud size, we would expect our result to be atomic
gas dominated. 

When averaged over individual clouds, the gas surface density is
higher by approximately a factor of 100, compared with averaging over
a larger area. In this volume, we can assume that the gas is at least
50\% molecular, depending on the mass of the assumed atomic
envelope. The fitted gradient for both these cloud
populations is $1.27$, slightly steeper than the result from
\citep{Bigiel2008}, likely due to this mix of
atomic and molecular gas. 

Note, that neither sets of points gets a gradient of 1.5, as would be
expected if our results were simply a product of having a constant
star formation efficiency term, as given in Equation~\ref{eq:starmass}.

Despite having a lower SFR over the first $2/3$\,rds of
the simulation, we do not see any great differences between the cloud
populations with and without diffuse heating. From
Figure~\ref{fig:sfr_history}, we can see that at our time of analysis,
t = 200\,Myr, the SFRs in both simulations are approximately
constant. However, the same plot at t = 150\,Myr shows no greater
disparity between them, due their difference in SFRs being accompanied
by similar differences in gas surface density, as was seen in
Section~\ref{sec:simtime}. There is slightly less
scatter in the simulation that includes the heating term, especially
when averaged over $750^2$\,pc$^2$, due to the less fragmented ISM
distribution. 

Focusing on the individual clouds, the specific star formation rate
(SSFR; the star formation rate per unit mass) can be plotted as a function of
the cloud's gas mass. In their recent paper, \citet{Lada2010} measure the
SFR in eleven GMCs in the Milky Way. They find a
linear dependence with the cloud's mass, specifically SSFR$ = 4.6 \pm
2.6 \times 10^{-8}$\,yr$^{-1}$. The relation for our clouds is plotted
in Figure~\ref{fig:ssfr} for the SSFR averaged over one dynamical
time. The left-hand plot shows the results for disk SFOnly and the
right-hand plot for disk SF+PEheat. Black circles show binned data for the complete cloud
sample while the red squares and blue triangles only include clouds
that have undergone a merger in the last dynamical time, with the
latter considering only major mergers with a cloud mass ratio less
than 2.0. 

The upper dotted line shows the value from \citet{Lada2010}. With the
exception of the most massive clouds, our star formation rate is lower
by a factor of 3-4. However, \citet{Lada2010} calculate the cloud mass
from gas above a threshold surface density of $\Sigma_{\rm gas}
\approx 116$\,M$_\odot$pc$^{-2}$, which they estimate is equivalent to
a volume density of $n_{\rm H_2} \approx 10^4$\,cm$^{-3}$. While our
clouds larger than $M \approx 10^5$\,M$_\odot$ achieve this surface
density, their volume density is much lower, with the majority of
clouds having average densities between $10^2 - 10^3$\,cm$^{-3}$. If
we reduce the \citet{Lada2010} SSFR value by a factor of 10 to allow
for this, we get the lower dotted line. Our cloud SSFR largely lie
above this lower line, suggesting that the SFR per cloud is
too high, given the resolution, by a factor of 10 in agreement with
Figure~\ref{fig:schmidt} for the disk averages over the same spatial scale
as the observations. 

For clouds with masses $M < 10^{5.5}$\,M$_\odot$, the SSFR is
approximately constant. At first sight, this seems to be in agreement
with the \citet{Lada2010} result, however this population of clouds are at
our resolution limit where their internal structure cannot
be resolved well. In our cloud population, density scales with mass so
larger objects have shorter dynamical times, increasing their SSFR. In
this region where our results are not resolution limited, $M \gtrsim
10^5$\,M$_\odot$, we therefore find that the SSFR is proportional
to the cloud mass. 

This result remains when we introduce diffuse heating in the
right-hand plot of Figure~\ref{fig:ssfr}. The maximum SSFR however is
less, in agreement with our previous findings, and there is evidence
of a down turn at high masses. This small population of very massive
($> 10^{6.5}$\,M$_\odot$) clouds are extended structures from recent
mergers, resulting in them having a lower density. 

As we saw in Figure~\ref{fig:mergers}, mergers between clouds are a
common occurrence and likely to have a significant impact on the clouds
evolution. \citet{Tan2000} suggested that such collisions could
trigger star formation, providing a way of connecting local-scale
motions with the globally observed Kennicutt-Schmidt relation. In
Figure~\ref{fig:ssfr}, however, we find that the presence of a recent
merger decreases the clouds SSFR. This effect is even greater if
the merger was a major one with a cloud whose mass was at least 50\%
of the object it merged with. The impact of the mergers depends on mass,
with larger clouds being unaffected unless experiencing a major merger. This is not hard to understand, since a more massive cloud will
be undergoing predominantly collisions with much smaller objects which
will cause less perturbations to its internal structure. The reason
for the drop in SSFR is due to a corresponding drop in the cloud
density. However, such a result is liable to be resolution dependent,
due to the difficulties in resolving the changes to the clouds'
internal structure across only a few cells. \citet{Federrath2010}
finds that 30 cells are required to resolve a vortex within a cloud, while local box simulations with resolutions on the sub-parsec scale
suggest that low density ($n_H = 3$\,cm$^{-3}$) colliding flows can
trigger local star formation by initiating gravitational collapse
\citep{Heitsch2008}. If this applies to higher density collisions,
then it is probable that cloud collisions could help, rather than
hinder, star formation. 

\section{Conclusions}

We investigated the formation and evolution of the GMC population
formed in two isolated, Milky Way-type galaxies at a resolution
$\lesssim 10$\,pc. Both our simulated disks included star formation and
radiative cooling, with our second model also including a diffuse
heating term, representative of the photoelectric heating from dust
grains. We did not include any form of localized energetic feedback
(e.g. supernovae). 

Both the disks fragmented through gravitational instabilities to form
a population of clouds, which we identified as GMCs when their
densities reached the threshold value of $n_{\rm H, c} >
100$\,cm$^{-3}$. These were tracked over time to provide statistics
both as a function of simulation time and of cloud age. 

The number of clouds formed above $M_c > 10^5$\,M$_\odot$ was found to agree well with the observed Milky Way
GMC population. The properties of the clouds were also comparable to
observations, including the distributions
of cloud mass, size, mass surface density, virial parameter, angular
momentum, vertical height above the disk and the distribution of
angles of angular momentum with respect to the galactic rotation
axis. 

Cloud ages were found to lie largely between $0-20$\,Myr, in good
agreement with current estimates. It is notable that this is without a
source of localized feedback which has previously been expected to provide
a dominant mechanism for cloud destruction. Many of our clouds die in
the first 3\,Myrs, reducing our population by 50\%. This cloud infant
mortality is due to mergers with nearby forming clouds and star
formation which can destroy low mass ($M < 10^5$\,M$_\odot$) objects.

The inclusion of diffuse heating raised the pressure of the warm and
cold ISM to suppress the fragmentation of the disk, producing an
initially smaller population of clouds embedded in
a more massive and structured warm ISM. The denser ISM retains a
filamentary structure after fragmentation that maintains a lower
velocity dispersion than when heating was absent. This environment has
two major impacts on the cloud properties: 

\begin{enumerate}
\item The filamentary warm ISM produces a predominantly
prograde rotating population of clouds even at late times. Without
diffuse heating, the clouds become $1/3$\,rd retrograde both with and
without star formation. This is evidence that the environment of the cloud
plays a dominant role in its evolution. 

\item The second effect is that the lower mass of cloud material in the disk
reduces the star formation rate in the first $175$\,Myr of
the simulation. Past this time, the star formation rate remained
approximately constant, while in the absence of heating, gas depletion
in the clouds causes the production of stars to drop by a factor of 10. 

\end{enumerate}

Cloud mergers and interactions were a frequent occurrence in both
disks, occurring at a rate of $\sim 0.25$ of an orbital period. This is
only slightly higher than the merger rate recorded in TT09 of the same
simulation without star formation. The effect of mergers appears to be
to reduce the cloud density, thereby reducing the star formation rate,
but we note this effect is likely to be dependent on resolution.

The star formation rate in the disks was roughly a factor of 10 too
high in both the disk without diffuse heating and when it was
included. There are several possibilities why this could be the case
including the absence of localized feedback and processes such as
added support from magnetic fields. Previous work that explores the
impact of magnetic fields on the GMC population suggests that the
magnetic pressure can suppress the formation of the
population. However, research performed by \citet{Dobbs2008a}, suggests
that the gas to magnetic pressure must be $\beta \lesssim 0.1$ for the
magnetic fields to suppress fragmentation in cold gas. Alone, the magnetic
Parker instability is not thought to be able to produce a realistic population
of GMCs, although it might play a role in seeding the gravitational
collapse \citep{Kim2006}. At the other end of the scale, localized
feedback from sources such as supernovae or radiation driven winds can
destroy clouds once they have formed stars. The exploration of both
these mechanisms will be explored in future studies. 

In addition to the physical processes not yet included, a second
source of error has to be the limit of our resolution. On average, our
clouds contain 76 cells and, for an average radius of 16\,pc, this
equates to roughly 4 cells in each dimension, although it is worth
noting that our average radius in the plane of the disk is larger, at
34\,pc. The impact this has on our calculations for the cloud
properties was investigated in TT09. Figure\,11 in that paper shows the
effect of reducing our limiting resolution to 15.6\,pc and 31.2\,pc,
plotting the cloud properties shown in Figures~\ref{fig:simtime_clouds_sf5} and
\ref{fig:simtime_clouds_peheat} for the populations in these runs. We
found that clouds with masses $M \gtrsim 10^6$\,M$_\odot$ were converged
at all resolutions, but the peak mass, radius and virial parameter
decreased with increasing refinement. Of particular concern was the
sensitivity of the rotational dynamics of the clouds to resolution. \citet{Federrath2010}
find that 30 cells are needed to accurately resolve a vortex, which is
almost a factor of ten more than the resolution we are currently able
to achieve within our clouds. To test the impact of this, we plotted
the variation of the clouds' angular momentum with respect to the
disk's rotation in Figure~\ref{fig:angmom_cloudt} for clouds with
masses greater than $M
> 10^6$\,M$_\odot$, but saw no change in the distribution of
$\theta$. This is a promising indication that our result will remain
true for a more highly resolved cloud structure, but we note that we
are not able to test this directly at the present time. \\

% acknowlegements

The author would like to thank Jonathan Tan, Ralph Pudritz and James
Wadsley for helpful discussions. EJT also acknowledges the University
of Florida High-Performance
Computing Center for providing computational resources and support and
use of the NCSA TeraGrid.


\begin{thebibliography}{}

\bibitem[Banerjee et al.(2009)]{Banerjee2009} Banerjee, R., 
V{\'a}zquez-Semadeni, E., Hennebelle, P., 
\& Klessen, R.~S.\ 2009, \mnras, 398, 1082 


\bibitem[Bertoldi 
\& McKee(1992)]{Bertoldi1992} Bertoldi, F., \& McKee, C.~F.\ 1992, \apj, 395, 140 

\bibitem[Bigiel et al.(2008)]{Bigiel2008} Bigiel, F., Leroy, A.,
  Walter, F., Brinks, E., de Blok, W.~J.~G., Madore, B., \& Thornley,
  M.~D.\ 2008, \aj, 136, 2846 

\bibitem[Binney \& Merrifield(1998)]{BM1998} Binney, J., \& Merrifield, M.\ 1998, Galactic astronomy, Princeton, NJ : Princeton University Press, 1998 

\bibitem[Blitz et al.(1990)]{Blitz1990} Blitz, L., Bazell, D., \& Desert, F.~X.\ 1990, \apjl, 352, L13 

\bibitem[Blitz 
\& Rosolowsky(2004)]{Blitz2004} Blitz, L., \& Rosolowsky, E.\ 2004, arXiv:astro-ph/0411520 

\bibitem[Bolatto et al.(2008)]{Bolatto2008} Bolatto, A., D., Leroy, A. K., Rosolowsky, E., Walter, F., Blitz, L. 2008, \apj, 686, 948

\bibitem[Boulares \& Cox(1990)]{Boulares1990} Boulares, A., \& Cox, D.~P.\ 1990, \apj, 365, 544 

\bibitem[Bryan(1999)]{Bryan1999}Bryan,G.L. Comp. Phys. and Eng. 1999, 1:2, p.

\bibitem[Bryan \& Norman(1997)]{Bryan1997} Bryan, G.~L.~\& Norman, M.~L.\ 1997, ASP Conf.~Ser.~123: Computational Astrophysics; 12th Kingston Meeting on Theoretical Astrophysics, 363

\bibitem[Cox(2005)]{Cox2005} Cox, D.~P.\ 2005, \araa, 43, 337 

\bibitem[Crutcher (2005)]{Crutcher2005} Crutcher, R. M. 2005, in {\it Massive star birth: A crossroads of Astrophysics}, IAU Symp. 227, ed. by Cesaroni, R., Felli, M., Churchwell, E., Walmsley, M., (Cambridge: CUP), pp.98


\bibitem[Dobbs(2008)]{Dobbs2008} Dobbs, C.~L.\ 2008, \mnras, 391, 
844 

\bibitem[Dobbs 
\& Price(2008)]{Dobbs2008a} Dobbs, C.~L., \& Price, D.~J.\ 2008, \mnras, 383, 497 


\bibitem[Draine(1978)]{Draine1978} Draine, B.~T.\ 1978, \apjs, 36, 595 

\bibitem[Federrath et al.(2008)]{Federrath2008} Federrath, C., 
Klessen, R.~S., \& Schmidt, W.\ 2008, \apjl, 688, L79 

\bibitem[Federrath et 
al.(2010)]{Federrath2010} Federrath, C., Roman-Duval, J., Klessen, R.~S., Schmidt, W., \& Mac Low, M.-M.\ 2010, \aap, 512, A81 

\bibitem[Fukui et al.(2008)]{Fukui2008}
Fukui, Y., Kawamura, A., Minamidani, T., Mizuno, Y., Kanai, Y. et al. 2008, \apjs, 178, 56

\bibitem[Fukui et al.(2009)]{Fukui2009} Fukui, Y., et al.\ 2009, 
\apj, 705, 144 

\bibitem[Gammie et al.(1991)]{Gammie1991} Gammie, C.~F., Ostriker, 
J.~P., \& Jog, C.~J.\ 1991, \apj, 378, 565 

\bibitem[Genzel et al.(2010)]{Genzel2010} Genzel, R., et al.\ 
2010, \mnras, 407, 2091 

\bibitem[Glover \& Mac Low(2007a)]{Glover2007a} Glover, S.~C.~O., \& Mac Low, M.-M.\ 2007, \apjs, 169, 239

\bibitem[Glover \& Mac Low(2007b)]{Glover2007b} Glover, S.~C.~O., \& Mac Low, M.-M.\ 2007, \apj, 659, 1317 

\bibitem[Glover et al.(2010)]{Glover2010} Glover, S.~C.~O., 
Federrath, C., Mac Low, M.-M., \& Klessen, R.~S.\ 2010, \mnras, 404, 2 

\bibitem[Habing(1968)]{Habing1968} Habing, H.~J.\ 1968, \bain, 19, 421 

\bibitem[Heitsch et al.(2008)]{Heitsch2008} Heitsch, F., Hartmann, 
L.~W., Slyz, A.~D., Devriendt, J.~E.~G., 
\& Burkert, A.\ 2008, \apj, 674, 316 

\bibitem[Heitsch et al.(2009)]{Heitsch2009} Heitsch, F., 
Ballesteros-Paredes, J., \& Hartmann, L.\ 2009, \apj, 704, 1735 

\bibitem[Hennebelle et 
al.(2008)]{Hennebelle2008} Hennebelle, P., Banerjee, R., V{\'a}zquez-Semadeni, E., Klessen, R.~S., \& Audit, E.\ 2008, \aap, 486, L43 

\bibitem[Joung \& Mac Low(2006)]{Joung2006} Joung, M.~K.~R., \& Mac Low, M.-M.\ 2006, \apj, 653, 1266 

\bibitem[Kennicutt(1998)]{Kennicutt1998} Kennicutt, R.~C., Jr.\ 1998, \apj, 498, 541 

\bibitem[Kim \& Ostriker(2001)]{Kim2001} Kim, W.-T., \& Ostriker, E.~C.\ 2001, \apj, 559, 70 

\bibitem[Kim et al.(2003)]{Kim2003} Kim, W.-T., Ostriker, E.~C., \& Stone, J.~M.\ 2003, \apj, 599, 1157 

\bibitem[Kim \& Ostriker(2006)]{Kim2006} Kim, W.-T., \& Ostriker, E.~C.\ 2006, \apj, 646, 213 

\bibitem[Kim \& Ostriker(2007)]{Kim2007} Kim, W.-T., \& Ostriker, E.~C.\ 2007, \apj, 660, 1232 

\bibitem[Kim et al.(2008)]{Kim2008} Kim, C.-G., Kim, W.-T., \& Ostriker, E.~C.\ 2008, \apj, 681, 1148 

\bibitem[Krumholz \& McKee(2005)]{Krumholz2005} Krumholz, M.~R., \& McKee, C.~F.\ 2005, \apj, 630, 250 

\bibitem[Krumholz \& Tan(2007)]{Krumholz2007} Krumholz, M.~R., \& Tan, J.~C.\ 2007, \apj, 654, 304

\bibitem[Kwan(1979)]{Kwan1979} Kwan, J.\ 1979, \apj, 229, 567 

\bibitem[Lada et al.(2010)]{Lada2010} Lada, C.~J., Lombardi, M., 
\& Alves, J.~F.\ 2010, arXiv:1009.2985

\bibitem[Lada \& Lada(2003)]{Lada2003} Lada, C. J., \& Lada, E. A. 2003, \araa, 41, 57

\bibitem[Larson(1981)]{Larson1981} Larson, R.~B.\ 1981, \mnras, 
194, 809 

\bibitem[McKee \& Ostriker(2007)]{McKee2007} McKee, C.~F., \& Ostriker, E.~C.\ 2007, \araa, 45, 565

\bibitem[McKee \& Ostriker(1977)]{McKee1977} McKee, C.~F., \& Ostriker, J.~P.\ 1977, \apj, 218, 148 

\bibitem[Murray \& Rahman(2010)]{Murray2010a} Murray, N., \& Rahman, M.\ 2010, \apj, 709, 424

\bibitem[Murray(2010)]{Murray2010} Murray, N.\ 2010, arXiv:1007.3270 

\bibitem[O'Shea et al.(2004)]{OShea2004} O'Shea, B.~W., Bryan, G., Bordner, J., Norman, M.~L., Abel, T., Harkness, R., \& Kritsuk, A.\ 2004, ArXiv Astrophysics e-prints, astro-ph/0403044

\bibitem[Robertson 
\& Kravtsov(2008)]{Robertson2008} Robertson, B.~E., \& Kravtsov, A.~V.\ 2008, \apj, 680, 1083 

\bibitem[Rosen \& Bregman(1995)]{Rosen1995} Rosen, A.~\& Bregman, J.~N.\ 1995, \apj, 440, 634

\bibitem[Rosolowsky et al.(2003)]{Rosolowsky2003} Rosolowsky, E., Engargiola, G., Plambeck, R., \& Blitz, L.\ 2003, \apj, 599, 258 

\bibitem[Sarazin \& White(1987)]{Sarazin1987} Sarazin, C.~L.~\& White, R.~E.\ 1987, \apj, 320, 32

\bibitem[Shetty 
\& Ostriker(2006)]{Shetty2006} Shetty, R., \& Ostriker, E.~C.\ 2006, \apj, 647, 997 

\bibitem[Shetty \& Ostriker(2008)]{Shetty2008} Shetty, R., \& Ostriker, E.~C.\ 2008, \apj, 684, 978 

\bibitem[Shetty et al.(2010)]{Shetty2010} Shetty, R., Glover, S.~C.,
  Dullemond, C.~P., \& Klessen, R.~S.\ 2010, arXiv:1011.2019 

\bibitem[Slyz et al.(2005)]{Slyz2005} Slyz, A.~D., Devriendt, J.~E.~G., Bryan, G., \& Silk, J.\ 2005, \mnras, 356, 737

\bibitem[Solomon et al.(1987)]{Solomon1987} Solomon, P.~M., Rivolo, 
A.~R., Barrett, J., \& Yahil, A.\ 1987, \apj, 319, 730 

\bibitem[Stark \& Lee(2005)]{Stark2005} Stark, A.~A., \& Lee, Y.\ 2005, \apjl, 619, L159 

\bibitem[Stone \& Norman(1992)]{Stone1992} Stone, J.~M.~\& Norman, M.~L.\ 1992, \apjs, 80, 753 

\bibitem[Tan(2000)]{Tan2000} Tan, J.~C.\ 2000, \apj, 536, 173

\bibitem[Tasker \& Bryan(2006)]{Tasker2006} Tasker, E.~J., \& Bryan, G.~L.\ 2006, \apj, 641, 878

\bibitem[Tasker \& Bryan(2008)]{Tasker2008} Tasker, E.~J., \& Bryan, G.~L.\ 2008, \apj, 673, 810 

\bibitem[Tasker et al.(2008)]{Tasker2008b} Tasker, E.~J., Brunino, 
R., Mitchell, N.~L., Michielsen, D., Hopton, S., Pearce, F.~R., Bryan, 
G.~L., \& Theuns, T.\ 2008, \mnras, 390, 1267 

\bibitem[Tasker \& Tan(2009)]{Tasker2009} Tasker, E.~J., \& Tan, J.~C.\ 2009, \apj, 700, 358 

\bibitem[Tonnesen 
\& Bryan(2010)]{Tonnesen2010} Tonnesen, S., \& Bryan, G.~L.\ 2010, \apj, 709, 1203 

\bibitem[Truelove et al.(1997)]{Truelove1997} Truelove, J.~K., Klein, R.~I., McKee, C.~F., Holliman, J.~H., Howell, L.~H., \& Greenough, J.~A.\ 1997, \apjl, 489, L179

\bibitem[Toomre(1964)]{Toomre1964} Toomre, A.\ 1964, \apj, 139, 1217 

\bibitem[Wada \& Norman(2007)]{Wada2007} Wada, K., \& Norman, C.~A.\ 2007, \apj, 660, 276

\bibitem[Williams \& McKee(1997)]{Williams1997} Williams, J.~P., \& McKee, C.~F.\ 1997, \apj, 476, 166 

\bibitem[Wong \& Blitz(2002)]{Wong2002} Wong, T., \& Blitz, L.\ 2002, \apj, 569, 157 


\bibitem[Wolfire et al.(1995)]{Wolfire1995} Wolfire, M.~G., 
Hollenbach, D., McKee, C.~F., Tielens, A.~G.~G.~M., 
\& Bakes, E.~L.~O.\ 1995, \apj, 443, 152 

\bibitem[Wolfire et al.(2003)]{Wolfire2003} Wolfire, M.~G., McKee, C.~F., Hollenbach, D., \& Tielens, A.~G.~G.~M.\ 2003, \apj, 587, 278

\bibitem[Zuckerman 
\& Evans(1974)]{Zuckerman1974} Zuckerman, B., \& Evans, N.~J., II 1974, \apjl, 192, L149 

\end{thebibliography}
\end{document}